\documentclass[3p]{elsarticle}
\makeatletter
\def\ps@pprintTitle{%
 \let\@oddhead\@empty
 \let\@evenhead\@empty
 \def\@oddfoot{\centerline{\thepage}}%
 \let\@evenfoot\@oddfoot}
\makeatother

\usepackage[english]{babel}
\usepackage[T1]{fontenc}
\usepackage[latin1]{inputenc}
\usepackage{amsmath,amsfonts,amsthm,amstext,amssymb}
\usepackage{tikz}
\usetikzlibrary{shapes.misc, positioning}
\usepackage{graphicx}
\usepackage{doi}
\usepackage{hyperref}
\usepackage{multicol}
\usepackage{enumitem}
\usepackage{nth}
\usepackage{siunitx}
\usepackage{makecell}

\usepackage{rotating}
\usepackage{caption}
\usepackage{subcaption}

\newcommand{\prt}[1]{\left( {#1} \right)}
\newcommand{\dv}[2]{\frac{\mathrm d{#1}}{\mathrm d{#2}}}

\begin{document}

\begin{frontmatter}

\title{COVID-19 Belgium: Extended SEIR-QD model with nursing homes and long-term scenarios-based forecasts}

\author{Nicolas Franco}
\ead{nicolas.franco@unamur.be}
\address{Namur Institute for Complex Systems (naXys) and Department of Mathematics, University of Namur, Namur, Belgium}
\address{Interuniversity Institute of Biostatistics and statistical Bioinformatics (I-BioStat) and Data Science Institute,
University of Hasselt, Hasselt, Belgium\vspace{-1cm}}

\begin{abstract}
Following the spread of the COVID-19 pandemic and pending the establishment of vaccination campaigns, several non pharmaceutical interventions such as partial and full lockdown, quarantine and measures of physical distancing have been imposed in order to reduce the spread of the disease and to lift the pressure on healthcare system. Mathematical models are important tools for estimating the impact of these interventions, for monitoring the current evolution of the epidemic at a national level and for estimating the potential long-term consequences of relaxation of measures. In this paper, we model the evolution of the COVID-19 epidemic in Belgium with a deterministic age-structured extended compartmental model. Our model takes special consideration for nursing homes which are modelled as separate entities from the general population in order to capture the specific delay and dynamics within these entities. The model integrates social contact data and is fitted on hospitalisations data (admission and discharge), on the daily number of COVID-19 deaths (with a distinction between general population and nursing home related deaths) and results from serological studies, with a sensitivity analysis based on a Bayesian approach. We present the situation as in November 2020 with the estimation of some characteristics of the COVID-19 deduced from the model. We also present several mid-term and long-term projections based on scenarios of reinforcement or relaxation of social contacts for different general sectors, with a lot of uncertainties remaining.
\end{abstract}


\end{frontmatter}
\vspace{-0.5cm}

\section{Introduction}

While there are many models addressing the COVID-19 pandemic, it is important to have models representing each specific country since the evolution of the outbreak as well as the mandated control measures and their efficacies are different. Compartmental SEIR-type epidemic models \cite{Rock:2014aa} --- where the population is divided into some compartments such as Susceptible, Exposed, Infectious and Recovered --- are very suitable for long term projections due to their potential computational speed of running different scenarios, in comparison to e.g.~individual-based models. Moreover, SEIR-QD variants --- with additional compartments concerning hospitalisations (Q because hospitalisation status is quite similar to quarantine) and deaths (D) --- are particularly well suited for COVID-19 pandemic due to the lack of unbiased information on the real prevalence \cite{Peng2020.02.16.20023465,Yang2020.03.12.20034595}.\\

We present one of the very few existing extended SEIR-QD model adapted and calibrated on Belgium situation and data. Two similar approaches have been developed by the SIMID COVID-19 team (UHasselt-UAntwerp) \cite{ABRAMS2021100449} and the BIOMATH team (UGent) \cite{Alleman2020.07.17.20156034}. Those independently developed models have their own characteristics and are complementary since it is difficult at this time to exactly know how to model COVID-19 in the best way. The main goal of those three models is to inform policymakers in Belgium about the  projections of potential future decisions as well as informing hospitals, institutions and the scientific community on the estimated effects of non pharmaceutical interventions (NPI). Alternative approaches have also been developed as an individual-based model \cite{Willem:2021tl} and a meta-population model \cite{Coletti2020.07.20.20157933}.\\

The three Belgian compartmental models have common characteristics as a calibration on hospitalisations, deaths and serological studies (but not considering cases data), a separation in several age classes with different characteristics, a distinction between asymptomatic, presymptomatic and symptomatic people with a different infectiousness, the use of social contact data \cite{Willem:2020aa} to monitor the transmission of the virus at different places (home, work, school, leisure) and a Bayesian sensitivity analysis using Markov Chain Monte Carlo (MCMC) methods. However, the model presented in this paper provides several improvements. The main one is the fact that nursing homes are modelled as isolated entities in order to account for differences in timing of spread of the coronavirus compared to the general population and for a proportion of non-COVID-19 related deaths in Belgian nursing homes collected data. Our model has no informed parameter (except social contact data) in order to recover different characteristics of COVID-19 and is calibrated on different stages of the hospitalisation path (admission, discharge and death) to get a good view on length of disease and hospital stay. There is also a specific estimation of potential reimportations coming from travellers during the holiday period to avoid an overestimation of the national transmission.\\

The paper is organised as follows. In Section \ref{secMM}, we present a technical description of the model. The main characteristics are presented in Subsection \ref{secdesc}, equations in Subsection \ref{seceq}, precisions on the data in Subsection \ref{secdata} and explanations of the calibration method and sensitivity analysis in Subsection \ref{seccal}. Additional details including the timeline used and the full set of estimated parameters of the model are given in \ref{secappend}. The Results and Discussion Section \ref{secRD} starts with a presentation of the current estimation from the model in  Subsection \ref{secesti} with different indicators including reproduction number and infection fatality rate at different periods as well as some characteristics of the COVID-19 disease. Then we present a test on the validity of the model in Subsection \ref{secprevcal} with the confrontation of more recent data with previous calibrations. In Subsection \ref{secmid}, we analyse a mid-term projections based on estimations of new policy measures applied in October and November in Belgium concerning hospitalisations and deaths together with an extrapolation on prevalence and seroprevalence within each age group. Then some scenarios-based long-term projections are presented in Subsection \ref{seclong} visualising potential impacts of various exit strategies during the first semester of 2021. Finally, in Subsection \ref{conclusion} we provide a conclusion with strengths and limitations concerning the presented model.

\section{Materials and Methods}\label{secMM}

\subsection{General description of the model}\label{secdesc}

\begin{figure}[!h]
\vspace{0.5cm}
\tikzstyle{fleche}=[->,>=latex, thick]
\tikzstyle{fleche2}=[->,>=latex, thick, dashed]
\tikzstyle{fleche3}=[->,>=latex, thick, dotted]
\scalebox{1}{
\begin{tikzpicture}
\node[above right] at (0,0) {\underline{General population (age classes $i=$ 0-24, 25-44, 45-64, 65-74, 75+):}};
\node (S) at  (0,-0.5) [draw, rounded rectangle,black,fill=green!40, align=center, text width=1.5cm] {\large $S_i$ {\tiny Susceptible}};
\node[below right= 40pt and -5pt of S] (E) [draw, rounded rectangle,black,fill=yellow!40, align=center, text width=1.3cm]  {\large $E_i$ {\tiny Exposed}};
\node[right= 30pt of E] (AI) [draw, rounded rectangle,black,fill=orange!40, align=center, text width=5.2cm]  {\large $\qquad\qquad\qquad  I^A_i \qquad\qquad\qquad  $ {\tiny Asymptomatic Infectious}};
\node[below right= 15pt and 60pt of E] (PI) [draw, rounded rectangle,black,fill=orange!40, align=center, text width=1.8cm]  {\large $I^P_i$ {\tiny Presymptomatic Infectious}};
\node[right= 20pt of PI] (SI) [draw, rounded rectangle,black,fill=red!40, align=center, text width=1.6cm]  {\large $I^S_i$ {\tiny Symptomatic Infectious}};
\node[below right= 20pt and 40pt of SI] (Q) [draw, rounded rectangle,black,fill=violet!40, align=center, text width=1.8cm]  {\large $Q_i$ {\tiny Quarantined =Hospitalised}};
\node[right= 30pt of Q] (D) [draw, rounded rectangle,black,fill=brown!40, align=center, text width=1.5cm]  {\large $D_i$ {\tiny Deceased}};
\node[right= 30pt of SI, above = 80pt of D] (R) [draw, rounded rectangle,black,fill=teal!40, align=center, text width=1.5cm]  {\large $R_i$ {\tiny Recovered}};
\draw[fleche] (S)--(E) node[midway,right]{\small$\sum_j M_{ij} \prt{\lambda_a (I^A_j+I^P_j) + \lambda_s I^S_j}$};
\draw[fleche2]  plot [smooth,tension=1] coordinates {(S.east) (5.5,-1.7) (E.north east)};
\node[right] at  (5.5,-1.7) {\it\small reimportations from travellers};
\draw[fleche] (E)--(AI) node[midway,above]{$\sigma.{p_a}_i $};
\draw[fleche] (E)--(PI) node[midway,left]{$\sigma.(1-{p_a}_i)\ $};
\draw[fleche] (PI)--(SI) node[midway,below]{$\tau$};
\draw[fleche] (SI)--(Q) node[midway,left]{$\delta_i$};
\draw[fleche] (AI.east)--(R) node[midway,above]{${\gamma_a}_i$};
\draw[fleche] (SI)--(R) node[midway,above]{${\gamma_s}_i$};
\draw[fleche] (Q)--(R) node[midway,left]{${\gamma_q}_i(t)\ $};
\draw[fleche] (Q)--(D)node[midway,below]{$r_i(t)$};
\node[above right] at (0,-7.5) {\underline{nursing homes (2000 separated copies):}};
\node (Sh) at  (0,-8) [draw, rounded rectangle,black,fill=green!40, align=center, text width=1.5cm] {\large $S_h$ {\tiny Susceptible}};
\node[below right= 40pt and -5pt of Sh] (Eh) [draw, rounded rectangle,black,fill=yellow!40, align=center, text width=1.3cm]  {\large $E_h$ {\tiny Exposed}};
\node[right= 30pt of Eh] (AIh) [draw, rounded rectangle,black,fill=orange!40, align=center, text width=5.2cm]  {\large $\qquad\qquad\qquad  I^A_h \qquad\qquad\qquad  $ {\tiny Asymptomatic Infectious}};
\node[below right= 15pt and 60pt of Eh] (PIh) [draw, rounded rectangle,black,fill=orange!40, align=center, text width=1.8cm]  {\large $I^P_h$ {\tiny Presymptomatic Infectious}};
\node[right= 20pt of PIh] (SIh) [draw, rounded rectangle,black,fill=red!40, align=center, text width=1.6cm]  {\large $I^S_h$ {\tiny Symptomatic Infectious}};
\node[below right= 20pt and 40pt of SIh] (Qh) [draw, rounded rectangle,black,fill=violet!40, align=center, text width=1.8cm]  {\large $Q_h$ {\tiny Quarantined =Hospitalised}};
\node[above right= 0pt and 70pt of Qh] (Dh) [draw, rounded rectangle,black,fill=brown!40, align=center, text width=1.7cm]  {\large $D_{75+}$ {\tiny Deceased from hospitals}};
\node[below right= 0pt and 70pt of Qh] (Dh2) [draw, rounded rectangle,black,fill=brown!40, align=center, text width=1.6cm]  {\large $D_h$ {\tiny Deceased from homes}};
\node[right= 30pt of SIh, above = 40pt of Dh] (Rh) [draw, rounded rectangle,black,fill=teal!40, align=center, text width=1.5cm]  {\large $R_h$ {\tiny Recovered}};
\draw[fleche] (Sh)--(Eh) node[midway,right]{\small$ m_h \prt{\lambda_a (I^A_h+I^P_h) + \lambda_s I^S_h}$};
\draw[fleche2]  plot [smooth,tension=1] coordinates {(Sh.east) (5.5,-9.2) (Eh.north east)};
\node[right] at  (5.5,-9.2) {\it\small transmissions from visits};
\draw[fleche] (Eh)--(AIh) node[midway,above]{$\sigma.{p_a}_h $};
\draw[fleche] (Eh)--(PIh) node[midway,left]{$\sigma.(1-{p_a}_h)\ $};
\draw[fleche] (PIh)--(SIh) node[midway,below]{$\tau$};
\draw[fleche] (SIh)--(Qh) node[midway,left]{$\delta_h(t)$};
\draw[fleche] (AIh.east)--(Rh) node[midway,above]{${\gamma_a}_h$};
\draw[fleche] (SIh)--(Rh) node[midway,above]{${\gamma_s}_h$};
\draw[fleche] (Qh)--(Rh) node[midway,left]{${\gamma_q}_h(t)\ $};
\draw[fleche] (Qh)--(Dh)node[midway,below]{$r_h(t)$};
\draw[fleche] (SIh.south west)to[out=-120,in=180] node[midway,left]{$P_{cor}\tilde r_h(t)\quad$} (Dh2.west) ;
\draw[fleche3] (Sh.south west)to[out=-90,in=180,text width=4.8cm] node[midway,left]{\small\it \quad\quad non COVID-19 deaths \quad\quad officially reported as COVID-19} (Dh2.south west) ;
\draw[fleche3] (Rh.south west)to[out=-120,in=180,text width=4.4cm]  (Dh2.south west) ;
\draw[fleche2] (S.south west)--(Sh.north west)node[near end,right]{\it\small new arrivals from $S_{75+}$};
\end{tikzpicture}}
\caption{Schematic view of the compartmental model. Straight lines represent the usual flows of individuals for a SEIR-QD-type model. Susceptible individuals $(S)$ move to an exposed state $(E)$ and after a latent period either to a completely asymptomatic disease $(I^A)$ or to a path presymptomatic-symptomatic $(I^P \rightarrow I^S)$. They all recover $(R)$ except a portion of symptomatic ones who require hospitalisation $(Q)$ and either recover $(R)$ or die $(D)$. A significative proportion of symptomatic individuals in nursing homes directly die without passing through hospital $(D_h)$ (this effect is minimal within the general population and ignored here). All those straight line flows are considered continuous and proportional to the size of the initial compartment. In order to take account of an overreporting in Belgium data concerning individuals dying directly from nursing homes, an adjustment is performed with the dotted line flow.  Dashed lines represent specific flows which are discrete in time (performed each day) and represent either infections due to external transmissions (due to travels for the general population or to visits for nursing homes) or new arrivals to nursing homes in order to compensate deaths. Those specific flows are detailed in Subsection \ref{seceq} and parameters in Table \ref{table_par}.}
\label{schema}
\vspace{1cm}
\end{figure}
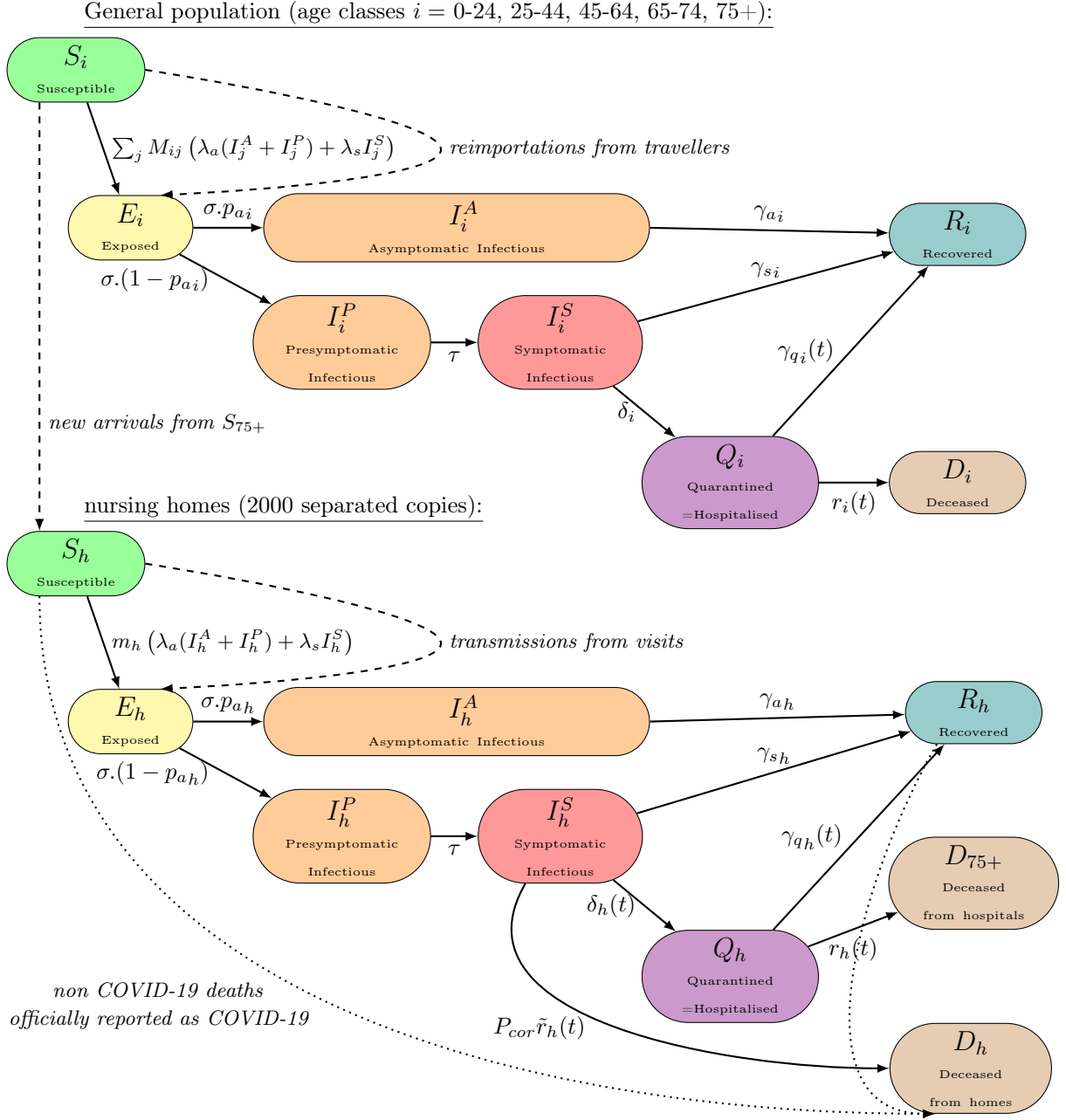

The continuous deterministic compartmental model is divided into the following 8 compartments in order to take account of the different possible stages of the disease as well as the separation between asymptomatic and symptomatic people with different infectiousness: Susceptible $S$, Exposed $E$, Asymptomatic Infectious $I^A$, Presymptomatic Infectious $I^P$, Symptomatic Infectious $I^S$, Hospitalised $Q$, Deceased $D$ and Recovered $R$. A more precise description is presented in Table \ref{compartments} of  \ref{secappend}. All those compartments exist for every age class. We do not consider in this model any subdivision inside the hospital compartment. A schematic view of the compartments with their relations is presented in Figure \ref{schema}.\\

 In addition, 2000 isolated nursing homes \cite{MRMRS} of similar average size are considered with all those compartments, also presented in Figure \ref{schema}, and modelled as isolated entities in order to take account of the different spread timing of the coronavirus compared to the general population. The transmission of infection from the general population to those nursing homes is modelled by a discrete random infection process which is detailed in Subsection \ref{sseceqnh}.  \\
 
We consider the following age classes among the population: 0-24, 25-44, 45-64, 65-74 and 75+. Those classes correspond to public available data \cite{SCIENSANO}. We assume that the classes up to 74 are only present among the general population, while the remaining is divided between a general 75+ and a specific class of nursing homes residents. The transmission of the coronavirus between all classes of the general population is computed using social contact data at different places \cite{Willem:2020aa}.\\

Some additional estimated parameters are considered in order to capture specific effects. A probability parameter is capturing the fact that only a part of the reported deaths from nursing homes are due to the COVID-19 \cite{NH}. A corrective coefficient is used to correct the new hospitalisations data since patients initially hospitalised for another reason or with no valid PCR test are not officially considered in the admissions data \cite{underreport}. Recovery and death rates from hospitals are considered variable in time in order to take the continuous improvement of care methods into account \cite{sciensanohosp}. A variable hospitalisation policy is considered for nursing homes during the first wave (period March-June) since residents are less likely to be hospitalised when the hospital load is important (more than half of the hospitals had admission criteria and specific agreements with nursing homes during the first wave \cite{KCE}). All those specificities are detailed in Subsection \ref{seceq}. This model takes into consideration potential reimportations of COVID-19 from abroad during the holidays period based on travel trends data which are detailed in Subsection \ref{secdata}. \\

Policy changes, according to Belgian epidemic' schedule, are modelled using different coefficients for the social contact matrices \cite{Willem:2020aa}. Social contacts are divided into 4 categories: home (household and nearby family), work (with transport), school and leisure (with other places). All contacts are considered at $100\%$ during the period up to March 14, 2020. Then reduced percentages are estimated by the model for the different periods of lockdown and phases of lift of measures. These reduced percentages are the effect at the same time of mobility restrictions, social distancing, prevention measures, testing and contact tracing, while it is mathematically impossible to determine the exact part of those effects. Hence new parameters for some or all social contact types are estimated each time there is an important policy change. The timeline of control measures in Belgium and the way those measures are modelled are described in \ref{secappend} and Table \ref{list_time}. Long-term scenarios-based projections are constructed assuming a constant policy and compliance to measures during the future with different realistic possibilities of percentage of social contacts for still unknown policy effects, but otherwise estimated impacts of previous control measures are assumed to remain the same in future.\\

This model does not take into consideration effects like seasonality or cross-immunity. The population is only age-structured and not spatially structured. A spatial refinement of such a model would be really important, but currently the public data officially provided are not detailed enough to properly fit a complex spatial-structured model. \\

\subsection{Equations of the model}\label{seceq}
 \subsubsection{Equations of the model for the general population part}

Equations of the model for the general population are the following ones, with $i=$ 0-24, 25-44, 45-64, 65-74, 75+ depending on the age class:
\begin{eqnarray*}
\dv{S_i}{t} &=& - S_i \sum_j M_{ij} \frac{\lambda_a (I^A_j+I^P_j) + \lambda_s I^S_j}{N_j}  \quad - \text{\it Infections during vacation travel }\\
&&- \text{\it Nursing homes new arrivals (for $S_{75+}$ only)}\\[0.2cm]
\dv{E_i}{t} &=& S_i \sum_j M_{ij} \frac{\lambda_a (I^A_j+I^P_j) + \lambda_s I^S_j}{N_j} - \sigma E_i \quad + \text{\it Infections during vacation travel }\\
\dv{I^A_i}{t} &=&  \sigma {p_a}_i E_i - {\gamma_a}_i I^A_i\\
\dv{I^P_i}{t} &=&  \sigma \prt{1-{p_a}_i} E_i - \tau I^P_i\\
\dv{I^S_i}{t} &=&  \tau I^P_i - \delta_i I^S_i -  {\gamma_s}_i I^S_i\\
\dv{Q_i}{t} &=&  \delta_i I^S_i - r_i(t) Q_i - {\gamma_q}_i(t) Q_i \\
\dv{D_i}{t} &=&   r_i(t) Q_i  \\
\dv{R_i}{t} &=& {\gamma_a}_i I^A_i +  {\gamma_s}_i I^S_i +  {\gamma_q}_i(t) Q_i  
\end{eqnarray*}

Compartments are Susceptible ($S$), Exposed ($E$), Infectious asymptomatic ($I^A_i$), presymptomatic ($I^P_i$) or symptomatic ($I^S_i$), Hospitalised ($Q$), Deceased ($D$) and Recovered ($R$). The main part of the model is continuous (time unit is days). Elements in italic are additional discrete actions performed each day. Infections during vacation travels are modelled as follows: during the holiday period (July-September 2020), some individuals are removed each day from the $S_i$ classes and added to the corresponding $E_i$ classes (for age classes below 75) according to estimated travellers and estimated prevalence in the visited countries (as explained in Subsection \ref{secdata}) with a global coefficient $C_\text{reimp}$.\\

Parameters are listed and explained in Table \ref{table_par} of \ref{secappend}. Some specific parameters are time-dependent and their dependence are computed using a logistic sigmoid function in order to model a smooth transition between two states with a minimal number of estimated parameters in order to minimise overfitting. For the general population, such a logistic function (called "recovery" function) monitors the continuous care improvement at hospitals over time \cite{sciensanohosp} (with parameters estimated from the data):
$$ {\gamma_q}_i(t) = {\gamma_q}_i \prt{ 1+ \frac{P_\text{recovery}}{1+e^{-\frac{t-\mu_\text{recovery}}{s_\text{recovery}}}}}  \qquad r_i(t) = r_i \prt{ 1- \frac{P_\text{recovery}}{1+e^{-\frac{t-\mu_\text{recovery}}{s_\text{recovery}}}}} $$

The structure of the population is $N_{0-24}=3250000$, $N_{25-44}=3000000$, $N_{45-64}=3080000$, $N_{65-74}=1150000$ and $N_{75+}=870000$ outside nursing homes (with an additional $N_{h}=150000$ inside nursing homes) for a total population of $N= 11500000$ (including death compartments) which is assumed constant.  Those numbers are round numbers coming from the structure of the Belgian population as provided by the Belgian Federal Government on April 2020 \cite{STATBEL}. The estimated initial prevalence $p_0$ is proportionally distributed between the $E_i$ on day 1 among the general population (corresponding to March 1 reported situation = February 29 real situation). Nursing homes are assumed not initially infected.\\

Transmission is governed by the so-called social contact hypothesis \cite{10.1093/aje/kwj317}. Social contact matrices $M_{ij}$ (representing the average number of contacts per day of age class $i$ from an individual of age classe $j$) are based on social contact data from Flanders (Belgium main region) collected in 2010 \cite{10.1371/journal.pone.0048695} and computed separately for home, work, school and leisure using the SOCRATES online tool \cite{Willem:2020aa}. Work and transport categories are merged as well as leisure and other places. Scaling coefficients $C_*$ are used to represent the differential effect of introduced lockdown/policy measures on each of these types of social contact. Those coefficients capture at the same time the transmission reduction coming from a global diminution of the contact rate (lockdown, closures) as well as from sanitary measures like social distancing or mask wearing. Hence the complete contact matrices are (for each given constant policy period detailed in \ref{secappend}):
$$ M_{ij} = C_\text{home} {M_{ij}}_\text{home} + C_\text{work} {M_{ij}}_\text{work} + C_\text{school} {M_{ij}}_\text{school} + C_\text{leisure} {M_{ij}}_\text{leisure}$$
In addition to the contact rate, there are two coefficients $\lambda_a$ and $\lambda_s$ representing the transmission probability for asymptomatic/presymptomatic and symptomatic individuals, capturing susceptibility and infectiousness. There are assumed age-class independent, while the heterogeneity in infectiousness is introduced by a distinct probability ${p_a}_i$ of being asymptomatic. \\

The basic reproduction number for the general population is estimated by the leading eigenvalue of the next-generation matrix \cite{Diekmann:1990aa,doi:10.1098/rsif.2009.0386} (with $i,j$ indexing the age-classes of the general population):
$$ R_0 = \text{maxeigenv}\left[  \lambda_a \prt{\frac{{p_a}_j}{{\gamma_a}_j} + \frac{1-{p_a}_j}{\tau}} M_{ij} + \lambda_s \prt{ \frac{1-{p_a}_j}{{\gamma_s}_j + \delta_j}} M_{ij} \right]_{ij}  $$
The effective reproduction number is estimated by $R_t = R_{0} \frac{\sum_i S_i(t)}{{\sum_i N_i}-\sum_i D_i(t)}$. Those reproduction numbers only capture the epidemic within the general population, while the situation within nursing homes is considered as a separated system (for which cases can arise due to contact with the general population through visits, but which does not itself affect the general population).
 
\subsubsection{Equations of the model for the nursing homes part}\label{sseceqnh}

Equations for the population in nursing homes are nearly similar to those of the general population:
\begin{eqnarray*}
\dv{S_h}{t} &=& - S_h m_h \frac{\lambda_a (I^A_h+I^P_h) + \lambda_s I^S_h}{75} \quad \prt{- \tilde r_h(t) (1-P_{cor}) I^S_h \ \text{\it if $S_h>0$}}\\
&&+ \text{\it New arrivals} - \text{\it Random transmissions from visits}\\[0.2cm]
\dv{E_h}{t} &=& S_h m_h \frac{\lambda_a (I^A_h+I^P_h) + \lambda_s I^S_h}{75}  - \sigma E_h + \text{\it Random transmissions from visits}\\
\dv{I^A_h}{t} &=&  \sigma {p_a}_h E_h - {\gamma_a}_h I^A_h\\
\dv{I^P_h}{t} &=&  \sigma \prt{1-{p_a}_h} E_h - \tau I^P_h\\
\dv{I^S_h}{t} &=&  \tau I^P_h - \delta_h I^S_h -  {\gamma_s}_h I^S_h  - \tilde r_h(t)  P_{cor} I^S_h\\
\dv{Q_h}{t} &=&  \delta_h I^S_h - r_h(t) Q_h - {\gamma_q}_h(t) Q_h \\
\dv{D_{75+}}{t} &+=&   r_h(t) Q_h  \\
\dv{D_h}{t} &=&   \tilde r_h(t)  I^S_h\  \\
\dv{R_h}{t} &=& {\gamma_a}_h I^A_h +  {\gamma_s}_h I^S_h +  {\gamma_q}_h(t) Q_h  \quad \prt{- \tilde r_h(t) (1-P_{cor}) I_h \ \text{\it if $S_h=0$}}
\end{eqnarray*}

Parameters for disease dynamics in nursing homes are distinct from those in the general population (except age-class independent ones, cf.~Table \ref{table_par}). There are 2000 nursing homes \cite{MRMRS} considered as separated entities, with a constant population of 75 inside each one, for a total of $N_{h}=150000$ residents (rounded up from official 2018 statistics \cite{healthybelgium}).  New arrivals are considered in order to maintain each nursing home's population and are removed from the 75+ susceptible class (while deaths originated from nursing home are considered as belonging to the general population, hence the nursing homes population remains constant). Those transfers from the general population $S_{75+}$ to nursing homes $S_h$ are taken into consideration since, according to the small population inside each nursing home, new arrivals can have a non-negligible effect on the proportion of susceptible residents. \\

Transmission inside each nursing home follows a usual SEIR-type transmission with a specific contact rate $m_h$. Transmissions coming from the general population is computed in a particular way using a daily probability of infection: each day, for each nursing home, one additional (integer) infected resident is moved from the $S_h$ compartment to the $E_h$ compartment with probability $P_{th}  S_h\sum_j  \frac{\lambda_a (I^A_j+I^P_j) + \lambda_s I^S_j}{N_j}$, where the coefficient distinguishes between the initial phase $P_{th}$ and lockdown and subsequent phases $P^\prime_{th}$. Note that this particular process is stochastic, as opposed to the rest of the model which is deterministic. Starting from lockdown, transmissions are only considered from the 25-65 population (i.e.~with $j=$ 25-44 and 45-64) since transmissions are mainly from nursing homes' workers. Potential reverse transmissions are however not monitored here i.e.~nursing home residents do not infect the general population because their impact is more negligible due to the huge size of the general population infected compartments.\\

 Deaths from nursing homes through hospitalisation are counted within the 75+ class for consistency with reported data.  Additional COVID-19 reported deaths from nursing homes (without hospitalisation) are monitored using an additional death rate $\tilde r_h$. Since the officially reported data combine both confirmed COVID-19 deaths and suspected COVID-19 deaths \cite{NH}, there is an unknown overreporting percentage within the data. This overreporting is captured by a constant probability $P_{cor}$ that deaths are COVID-19 related. Hence only $\tilde r_h(t)  P_{cor} I^S_h$ are removed from the symptomatic compartment while the remaining non-COVID-19 related deaths are assumed occurring in the susceptible class or in the recovered class if the first one is empty. For the first wave only (March 1 to June 30) a variable hospitalisation policy is computed (to represent the fact that the probability of hospitalisations of COVID-19 patients from nursing homes varied over time \cite{KCE}) using variable parameters $\delta_h(t)$ (proportion of hospitalised) and $\tilde r_h(t)$ (proportion of direct deaths) of constant sum $\delta_h(t)+P_{cor}\tilde r_h(t)=\delta_h$, this proportion being monitored over time by an estimated logistic function (called "hosp" function) depending on hospitals load with an additional delay:
$$ \delta_h(t) = \delta_h -  \frac{ \tilde r_h  P_\text{cor}}{1+e^{-\frac{Q(t-\text{delay})-\mu_\text{hosp}}{s_\text{hosp}}}} \qquad  \tilde r_h (t) =  \frac{ \tilde r_h }{1+e^{-\frac{Q(t-\text{delay})-\mu_\text{hosp}}{s_\text{hosp}}}}$$
This variable hospitalisation policy was not applied in the second wave since most of nursing home residents were hospitalised. Hence from July 1, those parameters are considered with the value $Q=0$.

\subsection{Data used}\label{secdata}

We consider the following data for the calibration of the model coming from Sciensano's (national public health institute of Belgium) public raw data \cite{SCIENSANO} (October 31, 2020 release), all in daily incidence: new hospitalisations, discharged and deaths from hospital, age-class specific deaths and deaths directly coming from nursing homes. Concerning new hospitalisations, an additional corrective estimated parameter SUPP$_{hosp}$ is added which estimates the percentage of missing COVID-19 patients at the time of admission (hence catching supplementary patients not initially hospitalised for COVID-19 or with no valid PCR test \cite{underreport}). This correction is directly applied to the data. Deaths reported with a specific date are considered on that specific date while situations reported by hospitals are considered to occur up to 24h before the hospital report hence 2 days before the official data communication. Note that graphics are plotted using the dates of Sciensano's communications (1 day delay).\\

Additional constraints are considered coming from Sciensano's epidemiological reports \cite{SCIENSANO}. Those constraints determine the set of admissibles parameters. Serological studies on blood donors during the first wave are considered to provide strong constraints on the prevalence. However, those serological data are biased since there are strict conditions to be blood donors: being between 18 and 75 years old and having not developed any COVID-19 symptoms during the previous weeks. This bias is naturally integrated into the model by considering for the fit the ratio between immune people coming directly from the asymptomatic compartment (hence the total number in $\sum_{i}  R_i$ coming from $I^A_i$ compartments, denote by $\sum_{i}  I^A_i\rightarrow R_i$) and the total asymptomatic population who has not developed a symptomatic COVID-19 disease ($\sum_{i}S_i+E_i+I^A_i+I^P_i+[I^A_i\rightarrow R_i] $) for the classes $i=$ 25-44, 45-64 and 65-74. This ratio should be respectively between $0.5\%$ and $2.8\%$ 7 days before March 30 and between $3.5\%$ and $6.2\%$ 7 days before April 14, April 27 and May 11 (the 7 day delay is here to take the needed time to build a detectable immunity into account).  There are also trivial constraints on parameters as e.g. ${\delta}_{0-24} < {\delta}_{25-44} < \dots$ in order to reproduce the increase severity of the COVID-19 for older people as well as trivial constraints to avoid negative or out-of-bound parameters.\\

 Additional constraints are imposed on nursing homes coming from the result of massive PCR test on April-May: the average percentage of infected people should be $8\% \pm 3\%$ during the period April 15-30  and less than $2\% \pm 2\%$ during the period May 15-31. Those percentages are estimated from Sciensano's epidemiological reports using a calculated incidence between each week. Additionally, the average percentage of asymptomatic residents (including presymptomatic ones) among infected should be $75\%\pm10\%$.\\
 
 This model takes into consideration potential reimportations of COVID-19 from abroad during the holiday period. No reimportation is assumed in June since borders where barely opened. Reimportations are estimated during the period July to September using the following method: According to 2019 travel trends \cite{ABTO}, we consider a proportionality of travellers of 36\% in July, 26\% in August and 21\% in September. There is no data available concerning the inhomogeneous repartition inside each month, but we assume a homogeneous one for July and August while a 2 to 1 ratio between the first half of September and the second half. Only the top five countries of destination are considered with the following proportion: France 23\%, Spain 11\%, Italy 9\% and The Netherlands 7\% (Belgium is discarded). Then for each of those countries we consider the daily ECDC statistics on cumulative numbers for the previous 14 days of COVID-19 cases per 100000 \cite{ECDC}. The reimportations are added using an estimated global coefficient $C_\text{reimp}$ and injected proportionally in the exposed compartment of the classes 0-24, 25-44, 45-64 and 65-74 and removed from the corresponding susceptible compartments. The estimated reimportations per day are presented in \ref{secappend} Table \ref{list_reimp}.

\subsection{Calibration method}\label{seccal}

Except social contact data, all of the 65 parameters of the model are estimated using a Markov Chain Monte Carlo (MCMC) method \cite{metropolis}, hence there is no assumption coming from studies in other countries. We assume that each daily incidence data follows a Poisson distribution which is appropriate when dealing with count data \cite{hilbe_2014}. The log-likelihood function, representing the probability that observed data correspond to model's projections, is computed as:
$$ \log L = \sum \prt{ -y_i \log(Y_i) } +  Y_i$$
where $y_i$ represent the observed incidences and $Y_i$ the expected incidences as given by the model for a given set of parameters. Note that the sum is done over all incidence data (new hospitalisations, discharged and deaths) presented in Subsection \ref{secdata} for each day and that a constant $\log(y_i !)$ is ignored.\\

The fitting procedure is performed in two steps:
\begin{itemize}\itemsep=0pt
\item Best-fit mode: An initial calibration step is performed using the maximum likelihood method with an optimised first-choice hill climbing algorithm performed half of the steps on one parameter at a time (i.e.~one neighbour = variation of one parameter) and the other half on all parameters (i.e.~one neighbour = variation of all parameters), with a quick best fit search performed on accepted descent directions to speed up the process. For all parameters, wide normal prior distributions are used (Table \ref{list_par}). This initial calibration is highly computationally demanding due to the presence of a very high number of estimated parameters and the presence of local minima. It is initially performed during 5000000 iterations with a special trick to increase the rapidity of the algorithm: instead of 2000 different nursing homes, only 100 nursing homes are considered with each time 20 copies of each. This approximation is suitable as long as the algorithm is still far from the best-fit. In a second time, the best-fit search is pursued for 20000 iterations using the complete 2000 different nursing homes in order to further refine selected parameters. All this procedure is repeated at least 1000 times using parallel computing and 250 parameter sets with best scoring model runs are retained (the others 75\% are discarded in order to avoid unwanted local minima).
\item MCMC mode: A classic Random-Walk Metropolis (RWM) algorithm \cite{metropolis,Lesaffre} is performed in order to provide Bayesian inference using the Poisson log-likelihood assumption with the algorithm initiated from the 250 parameter sets obtained from the best-fit mode. For each parameter set, a 20000 burn-in period is performed followed by 200000 iterations retaining every 20000th iteration, which provide 2500 final samples coming from potentially different local minima zones in order to avoid a too high autocorrelation of the results.
\end{itemize}
The program is written is C language. The code source is publicly available \cite{githubcorona}. The full ODEs are solved by numerical integration using the GNU gsl odeiv2 librairy and a Runge-Kutta-Fehlberg45 integrator. The computation was performed on the HPC cluster Hercules2 \footnote{"Plateforme Technologique de calcul Intensif" (PTCI) located at the University of Namur, Belgium.}.

\section{Results and Discussion}\label{secRD}
\subsection{Current estimations}\label{secesti}

\begin{figure}[!htb] 
\centering
\includegraphics[width=12.5cm]{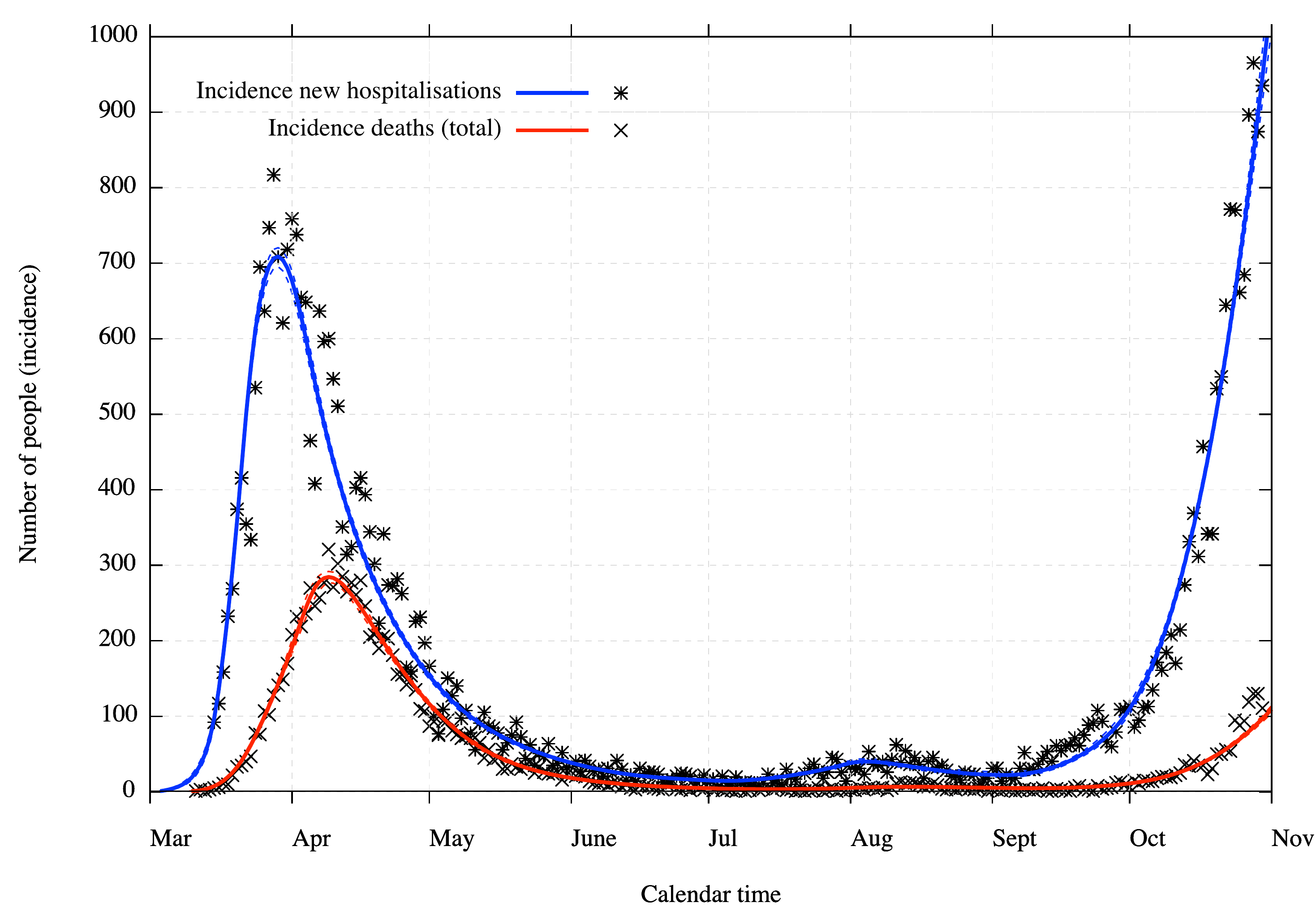}
\caption{Incidence in new hospitalisations (with underreporting correction included) and deaths with 90\% CI (dotted lines)}
\label{incid}
\end{figure}

\begin{figure}[!htb]  
\centering
\includegraphics[width=12.5cm]{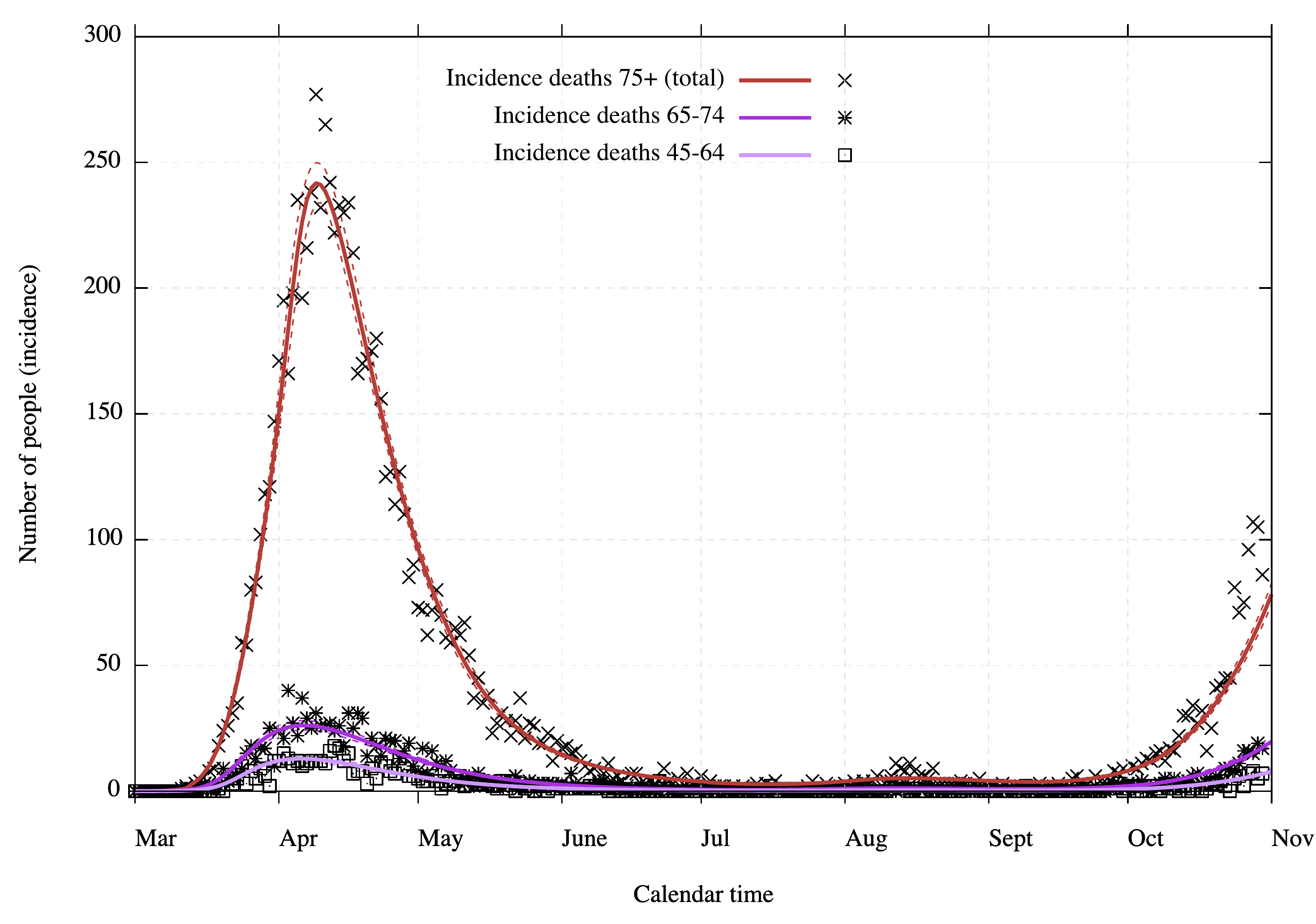}
\caption{Incidence deaths within age classes with 90\% CI (dotted lines)}
\label{ages}
\end{figure}

We present in this section the result of the calibration of the model as on November 1, 2020, with considered data up to October 31, 2020. Results are presented in the figures with medians, $5\%$ and $95\%$ percentiles, hence with a $90\%$ confidence interval. The comparison between the model and some hospitalisations and deaths incidence data are presented in Figure \ref{incid} for the general incidence data in hospitalisations and deaths and in Figure \ref{ages} for incidence data in deaths with age class repartition (for the classes which have a significative amount of deaths). Data in Figure \ref{incid} are plotted with the correction coming from the underreporting on new hospitalisations  to account for the discrepancies in COVID-19 hospital admissions, discharges and deaths. We can notice that, since this model is deterministic, it only captures the average evolution with an uncertainty concerning this average evolution, hence the uncertainty does not capture stochastic realisations around this average.

\begin{figure}[!htb] 
\centering
\includegraphics[width=15cm]{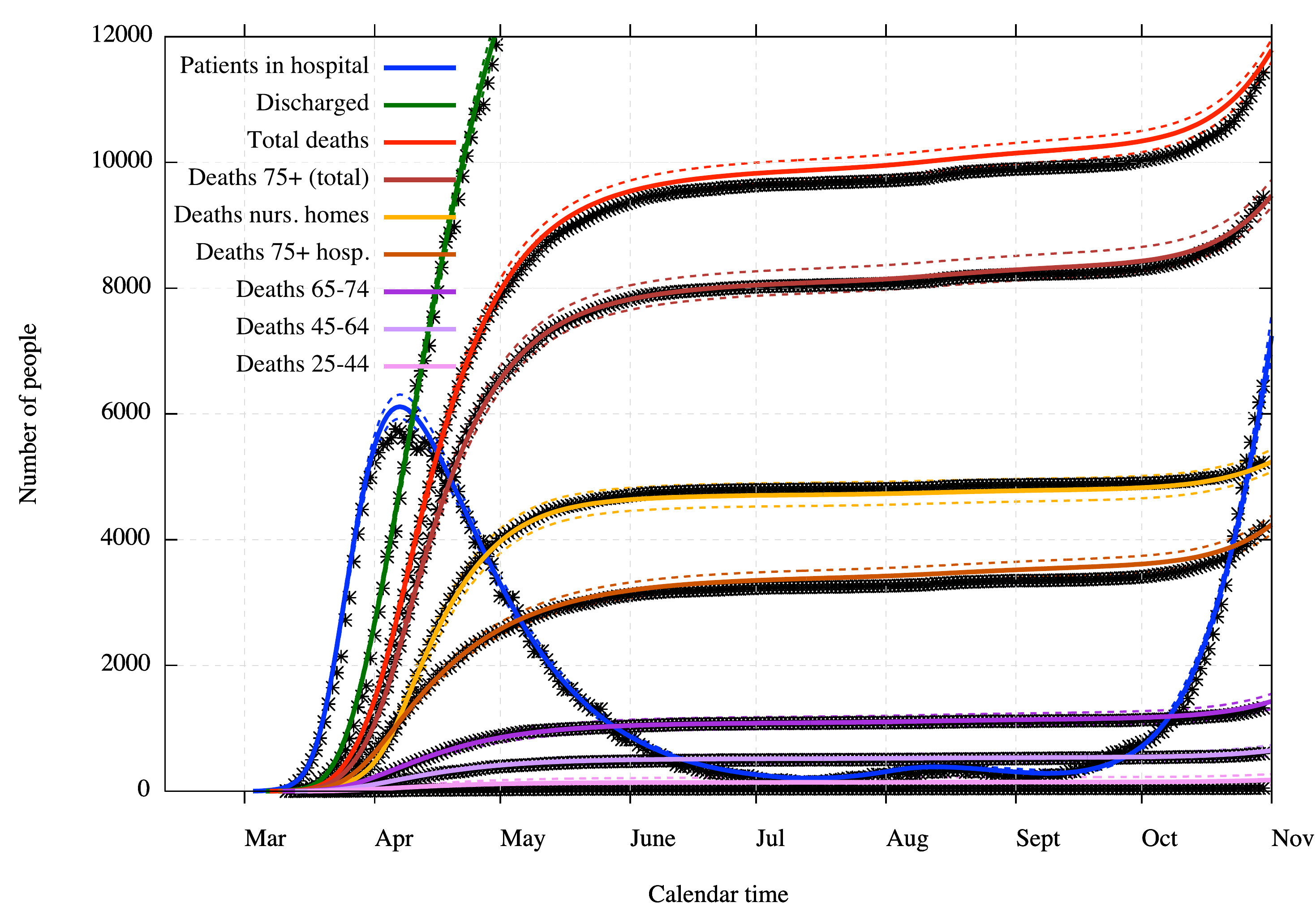}\vspace{-0.2cm}
\caption{General view on prevalence data and estimations with 90\% CI (dotted lines)}
\label{general}
\end{figure}

Figure \ref{general} shows a general representation of the evolution of the epidemic in Belgium with hospitalisations, people discharged from hospitals and deaths coming from hospitals and from nursing homes, all in prevalence or cumulative numbers. We can see that the model calibration fits the hospitalisation prevalence and cumulative deaths data with a good exactitude (excluding of course data noises) despite that fact that the calibration is entirely done on incidence data. The interest in modelling the epidemic within nursing homes separately from the general population can clearly be seen on this figure. Indeed, the form of the death curve for nursing homes is really different from the ones for the general population since the epidemic started later in nursing homes but was more significant.

Initially the model overestimated the number of deaths from the end of the first wave. It was not possible to calibrate constant death rates throughout all phases of the epidemic. This may be a consequence of either or both care improvement in hospitals and lower aggressiveness of the virus. Hence death and recovered rates within each age class are modified by a logistic function depending on time (cf.~Subsection \ref{seceq} for details). The current improvement (in comparison to the very beginning of the epidemic) is estimated as $58.2\%$ [$49.3\%$ ; $64.4\%$], hence $58.2\%$ of the patients which should have died in March are now recovering from hospitals. We remark that it is impossible to know which part is due to care improvement (which was confirmed \cite{sciensanohosp}) and which part is potentially due to lower aggressiveness of the virus (if there is any) and that the death rate seems to rise again in October.  However, Figure \ref{ages} presents a slightly larger increase than expected for the deaths within the class 75+. This could be indicative of a small decrease in the quality of care during the second wave due to the huge load of the hospitals (but still better than during the first wave).\\

Deaths coming directly from nursing homes are not all due to COVID-19 since many PCR tests are lacking. The model estimates that only $83.1\%$ [$66.9\%$ ; $89.4\%$] of direct nursing home deaths are really due to COVID-19. The ratio between deaths coming directly from nursing homes and deceased patients in hospitals coming initially from nursing homes seems to be not constant, and it was necessary to introduce a variable hospitalisation policy \cite{KCE}. The best answer found was to monitor hospitalisations from nursing homes through a logistic function depending on general hospital load but with a specific delay (cf.~description of the logistic function in Subsection \ref{sseceqnh}). Hence, when hospital load starts to become too high, less people from nursing homes are hospitalised and the reverse effect occurs when hospital load gets lower, but with a delay parameter estimated at 10.6 days [8.4 ; 12.9].\\

The basic reproduction number $R_0$, representing the average number of cases directly generated by one infectious case in a population which is assumed totally susceptible, is estimated in average for each period (we consider this number dependent on lockdown measures) and computed as the leading eigenvalue of the next-generation matrix (cf.~Subsection \ref{seceq} for details). The effective reproduction number $R_t$ represents the average number of cases directly generated by one infectious case taking account of the already immune population, hence varying over a period. Estimations for the general population are presented in Table \ref{table_r0}.

\begin{table}[ht!]
\centering\vspace{-0.5cm}
\footnotesize
\begin{tabular}{|c|c|c|}
\hline
 & $R_0$  & $R_t$ (at the end of the period)\\
 \hline
Pre-lockdown: March 1 $\rightarrow$ March 13 &  4.08 [3.90 ; 4.34] & 4.04 [3.86 ; 4.30] \\
School and leisure closed: March 14 $\rightarrow$ March 18 & 2.22 [2.13 ; 2.34] & 2.16 [2.07 ; 2.27] \\
Full lockdown: March 19 $\rightarrow$ May 3 & 0.65 [0.60 ; 0.71] & 0.61 [0.56 ; 0.66] \\
Phase 1-2: May 4 $\rightarrow$ June 7 & 0.79 [0.74 ; 0.83] & 0.73 [0.69 ; 0.78] \\
Phase 3: June 8 $\rightarrow$ June 30  & 0.98 [0.91 ; 1.06] & 0.91 [0.84 ; 0.98] \\
Phase 4: July 1 $\rightarrow$ June 28  & 1.37 [1.28 ; 1.50] & 1.27 [1.19 ; 1.39] \\
Phase 4bis: July 29 $\rightarrow$ August 31  & 0.73 [0.63 ; 0.88] & 0.68 [0.58 ; 0.82] \\
Second wave: September 1 $\rightarrow$ October 31  & 1.70 [1.62 ; 1.80] & 1.34 [1.28 ; 1.41] \\
 \hline
 \end{tabular}
 \caption{Estimations of $R_0$ and $R_t$ values}
 \label{table_r0}
 \end{table}
 
 The reproduction number of the pre-lockdown period is a bit overestimated compared to other Belgian models (\cite{ABRAMS2021100449},\cite{Willem:2021tl}, but in accordance with \cite{Alleman2020.07.17.20156034}). This is probably due to the fact that the model does not take explicitly account of infections coming from foreign travel at this particular time and this results in an estimated $R_0$ slightly above 4. For the period  May 4-June 7 phases 1A-1B-2 (cf.~Table \ref{list_time}), since there were policy changes almost every week, we only provide here the estimated $R_0$ at the end of this period. The second wave $R_0$ does not take account of the new measures applied in October 19 whose effects should only be visible on November.\\

 The infection fatality rate (IFR) can be estimated using the total set of recovered people according to the model (hence including untested and asymptomatic people). Due to variable death rates over time, the IFR in the early period of the epidemic is higher than in the later months. Estimations are presented in Table \ref{table_IFR}. The mean and last period are limited to September since October data need some consolidation regarding the number of deaths.
  
\begin{table}[ht!]
\centering
\footnotesize
\begin{tabular}{|c|c|c|c|}
\hline
 & General IFR & March-April period & July-September period \\
 \hline
Overall population & 1.04\% [0.93\% ; 1.14\%]  & 1.15\% [1.02\% ; 1.28\%] & 0.34\% [0.31\% ; 0.36\%] \\
0-24  & 0.01\% [0.00\% ; 0.02\%]  & 0.01\% [0.00\% ; 0.02\%]  & 0.00\% [0.00\% ; 0.01\%] \\
25-44 & 0.05\% [0.03\% ; 0.07\%]  &  0.06\% [0.04\% ; 0.07\%]  & 0.02\% [0.01\% ; 0.03\%]\\
45-64  & 0.21\% [0.20\% ; 0.22\%]  & 0.22\% [0.21\% ; 0.23\%]  &  0.09\% [0.08\% ; 0.10\%]\\
65-74 & 1.85\% [1.78\% ; 1.92\%] & 1.97\% [1.90\% ; 2.05\%]  & 0.97\% [0.93\% ; 1.02\%]  \\
75+ (nursing homes included)  & 8.34\% [7.57\% ; 9.36\%] & 9.75\% [8.81\% ; 10.99\%]  &  2.19\% [1.97\% ; 2.47\%]\\
 \hline
 \end{tabular}
 \caption{Infection fatality rate estimations}
 \label{table_IFR}
 \end{table}

Table \ref{table_carac} presents some model estimates characterising disease progression. Durations are derived according to some specific rate parameters related to the model. The model cannot really detect the exact time when symptoms appear, hence the end of the incubation period merely corresponds to the estimated time when the infectiousness becomes more important.  The total disease duration for symptomatic people concerns only not hospitalised individuals (not directly recovering from the $I^S_i$ compartment), while the duration is longer for the others. The hospitalisation duration is the average until discharged or deceased (no distinction is provided, hence according to the average rate of exit of the $Q_i$ compartment) at the beginning of the epidemic, hence before care improvement.  The duration for asymptomatic nursing homes' residents cannot really be estimated by the model (the confidence interval is very wide). Indeed, once a single nursing home is completely infected, asymptomatic infected residents can remain a very long time inside the $I^A_h$ compartment without infecting any new resident, hence there is no constraint within the model on this duration coming from the available data. This excessive duration must be considered as an outlier.

\begin{table}[ht!]
\centering
\footnotesize
\begin{tabular}{|c|c|c|c|c|c|c|}
\hline
&  0-24& 25-44& 45-64&65-74&75+&nursing homes\\
 \hline
Latent (pre-infectious) period & \multicolumn{6}{c|}{1.4 days [1.1 ; 2.0]} \\  \hline
Presymptomatic period & \multicolumn{6}{c|}{ 6.7 days [4.7 ; 8.0] }\\  \hline
Total disease duration   & 4.7 days & 5.2 days& 5.7 days & 6.3 days  & 7.8 days  & 27.3 days   \\  
asymptomatic people  & [4.1 ; 5.4]  &  [4.5 ; 6.0]&  [4.7 ; 6.5]   &  [5.5 ; 7.5]   &  [6.3 ; 10.0]   &  [17.0 ; 62.9] \\  \hline
Total disease duration    &11.2 days& 11.6 days & 12.1 days  & 12.7 days & 13.2 day & 13.9 days   \\  
symptomatic people  &[9.6 ; 12.4]  &  [10.3 ; 13.1] & [10.8 ; 13.6]  & [11.3 ; 14.0]  & [11.6 ; 14.6]  &[12.3 ; 15.3]  \\  \hline
Hospitalisation duration   & 15.4 days & 17.4 days& 16.4 days & 12.1 days  & 11.4 days  & 10.7 days   \\  
(before care improvement)  & [12.6 ; 17.9]  & [16.1 ; 18.9]  &  [15.2 ; 17.6]   & [11.1 ; 13.5] & [10.6 ; 12.5]  &  [9.0 ; 11.9]   \\  \hline
Overall percentage of    & 91.5\% & 84.3\% & 72.8\% & 55.8\% & 35.3\% & 25.7\%  \\
asymptomatic  people  & [78.4 ; 95.3]  &  [70.5 ; 90.1]  &  [60.3 ; 81.2]  &  [41.9 ; 64.8]  &  [23.9 ; 50.1]  &  [12.5 ; 38.5]  \\  \hline
 \end{tabular}
 \caption{Some characteristics of the COVID-19 as estimated by the model. All durations are average durations, and the given uncertainties are uncertainties on those averages, not on the individual values.}
 \label{table_carac}
 \end{table}
 
\subsection{Confrontation of previous calibrations}\label{secprevcal}
One way to test the robustness of a model is to assess ability of the model to predict the evolution of the epidemic beyond the time period for which data is used to fit the model. This model can provide projections or scenarios in two different ways. When new policy interventions are expected or a specific behaviour change is planned due to the calendar, it is possible to extrapolate the future transmission of the COVID-19  (monitored here by the number of contacts) using relative percentage of transmission in comparison to the pre-lockdown phase. This percentage can only be a vague estimate of what could be the real transmission and it is sometimes necessary to look at several different scenarios. On the other hand, when no new policy intervention is expected for a certain time, it is possible to have a more precise projection based on current estimated contacts (what we call the current behaviour) but which is only valid up to the next policy intervention.\\

\begin{figure}[!htb]  
\begin{subfigure}{.49\textwidth}
\centering
\includegraphics[width=\textwidth]{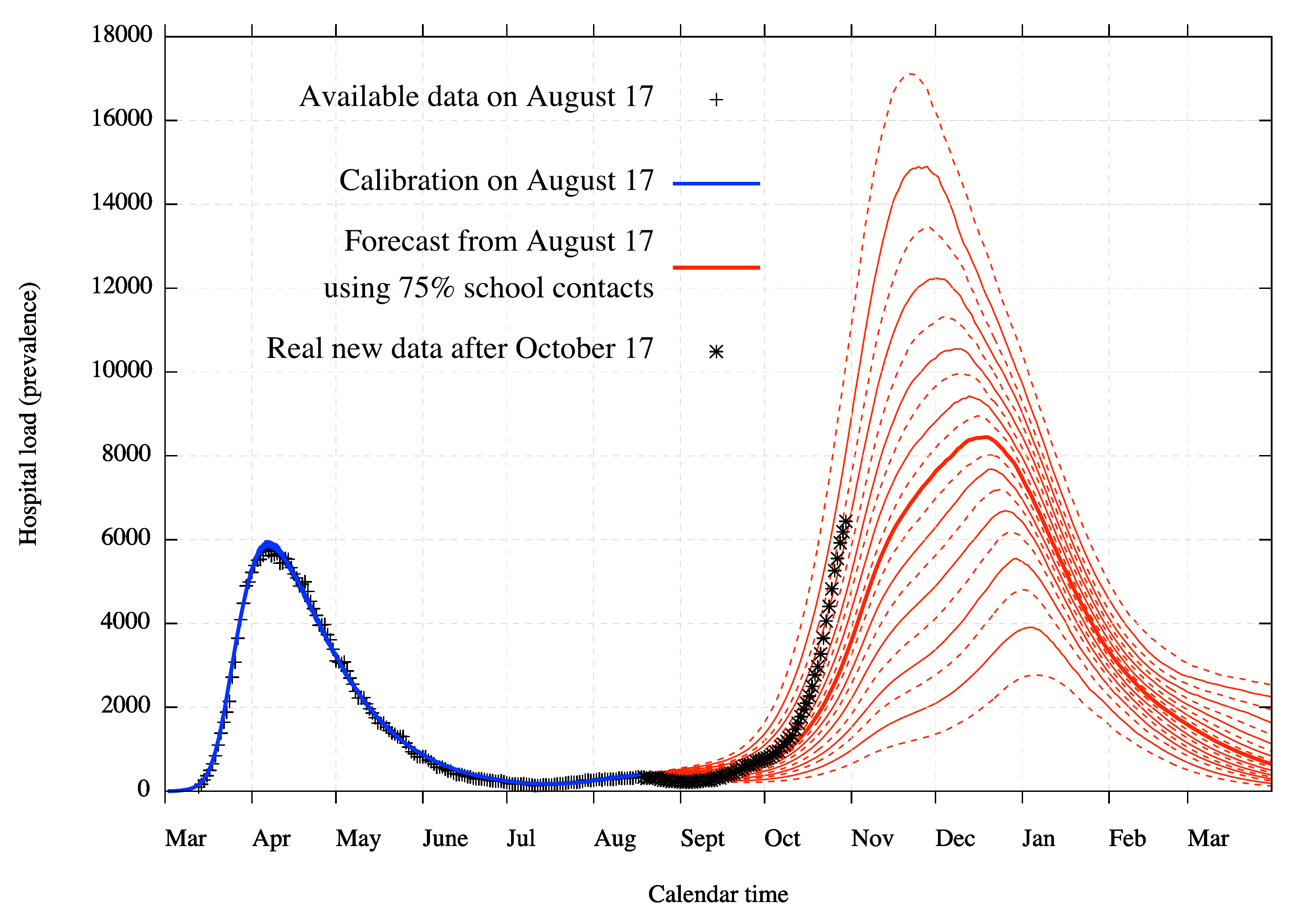}
\caption{Projection from August 17}
\label{prev_for_august}
\end{subfigure}
\begin{subfigure}{.49\textwidth}
\centering
\includegraphics[width=\textwidth]{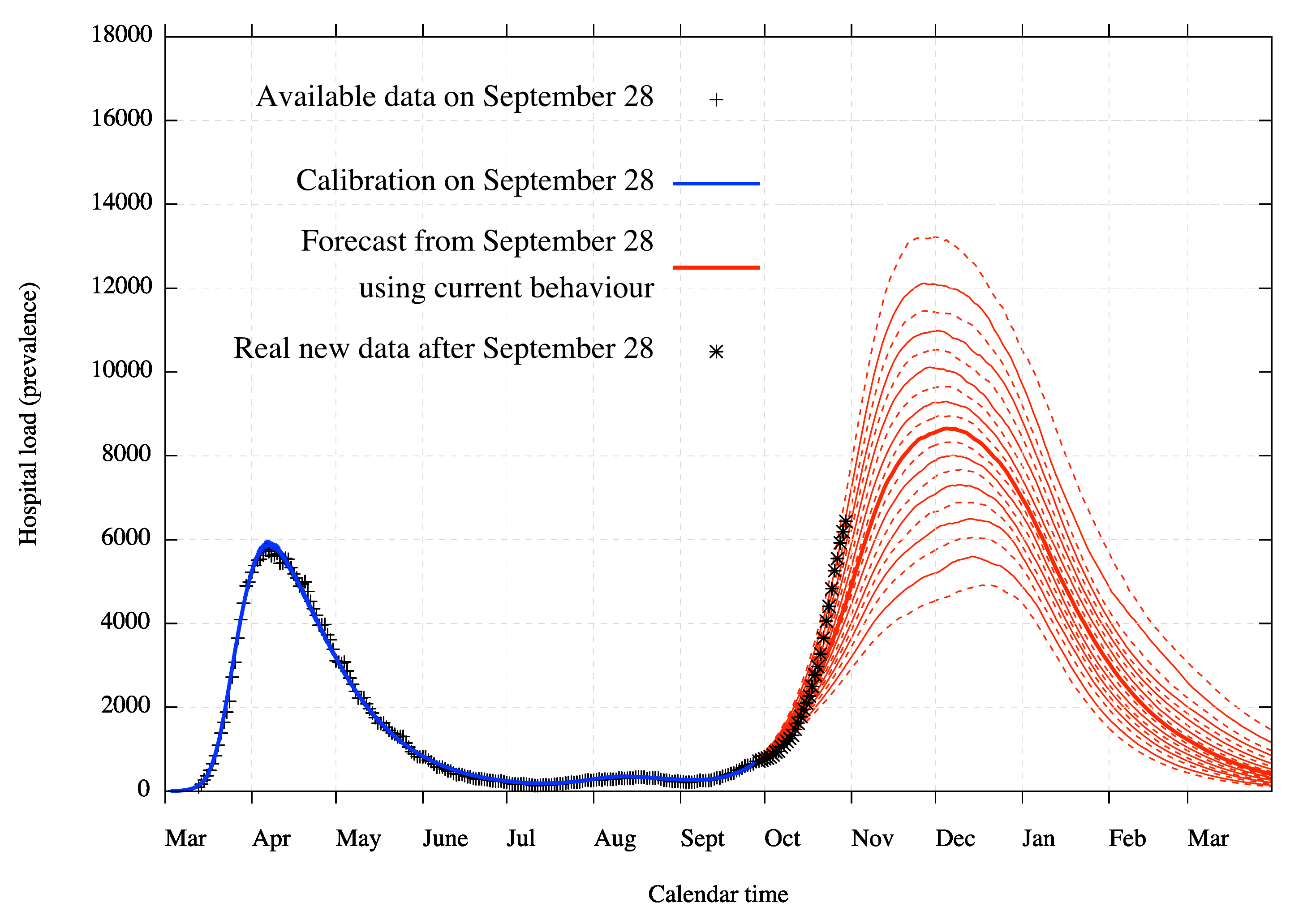}
\caption{Projection from September 28 }
\label{prev_for_sept}
\end{subfigure}
\caption{Previous projection from August 17 based on the scenario of a 75\% transmission at school from September 1 and  from September 28 based on the continuation of current behaviour. The strong line represents the median, continuous lines represent deciles (10\% percentiles) while dashed lines represent ventiles (5\% percentiles). Uncertainty covers both uncertainty about disease parameters and the impact of control measures.}
\label{fig_prev}
\end{figure}

We present two old projections from the model. The first one in Figure \ref{prev_for_august} is a 2.5 month projection based on the specific scenario that the transmission at school from September 1 would be at a level of 75\%  in comparison to the pre-lockdown period due to sanitary measures like masks wearing. The second one presented in Figure \ref{prev_for_sept} is a 1 month projection based on the current behaviour and the estimation from the model of the percentage of transmission at schools compared to the pre-pandemic situation (coefficient $C_\text{school}$), which was estimated at that time to be 69.7\% [44.2\% ; 88.6\%].  Comparison of these old projections with observations highlights the fact that the uncertainty must be taken into consideration.

\subsection{Mid-term scenario-based projection}\label{secmid}

Every scenario is hypothetical. New measures that have not been tested cannot really be estimated on the level of their impact and it is impossible to predict evolution in compliance to them from the population as well as future policy changes. This is why any realistic projection must rely on the assumption of a perfect continuity of measures and compliance for elements which are a priori not suspected to change soon and on different hypothetical scenarios for untested modifications of measures.\\

In response to the large second wave in Belgium, authorities decided to enforce new measures on October 19 such as closing bars and restaurants, reducing the allowed social contact (known as bubble) to one person, promoting teleworking and establishing a curfew night time. On November 2, a soft lockdown was put into place, with closure of non-essential shops, teleworking mandatory, leisure mostly reduced and social contacts even more reduced. Schools are closed during 2 weeks and then reopen with a 5/6 attendance (except for universities).\\

While it is impossible to know with precision the impact from those measures, we estimate that the effect from the soft lockdown could be comparable to the effect of the first lockdown, since the small remaining liberties could be balanced by generalised sanitary measures like mask wearing. The effects from October 19 measures are more uncertain but should be situated in terms of efficacy between September behaviour and lockdown behaviour. Hence the most realistic mid-term scenario is to consider a medium situation from October 19, with a full reduction applied from November 2 until the December 13 planned deadline. Schools are considered at 0\% transmission from November 2 to November 15 and at 5/6 thereafter. Every contacts are assumed to be restored at September level after December 13 (except for usual school closures). Social contact matrix coefficients concerning this scenario are detailed in Table \ref{list_time_sce}.\\

\begin{figure}[htb!]  
\begin{subfigure}{.5\textwidth}
\centering
\includegraphics[width=\textwidth]{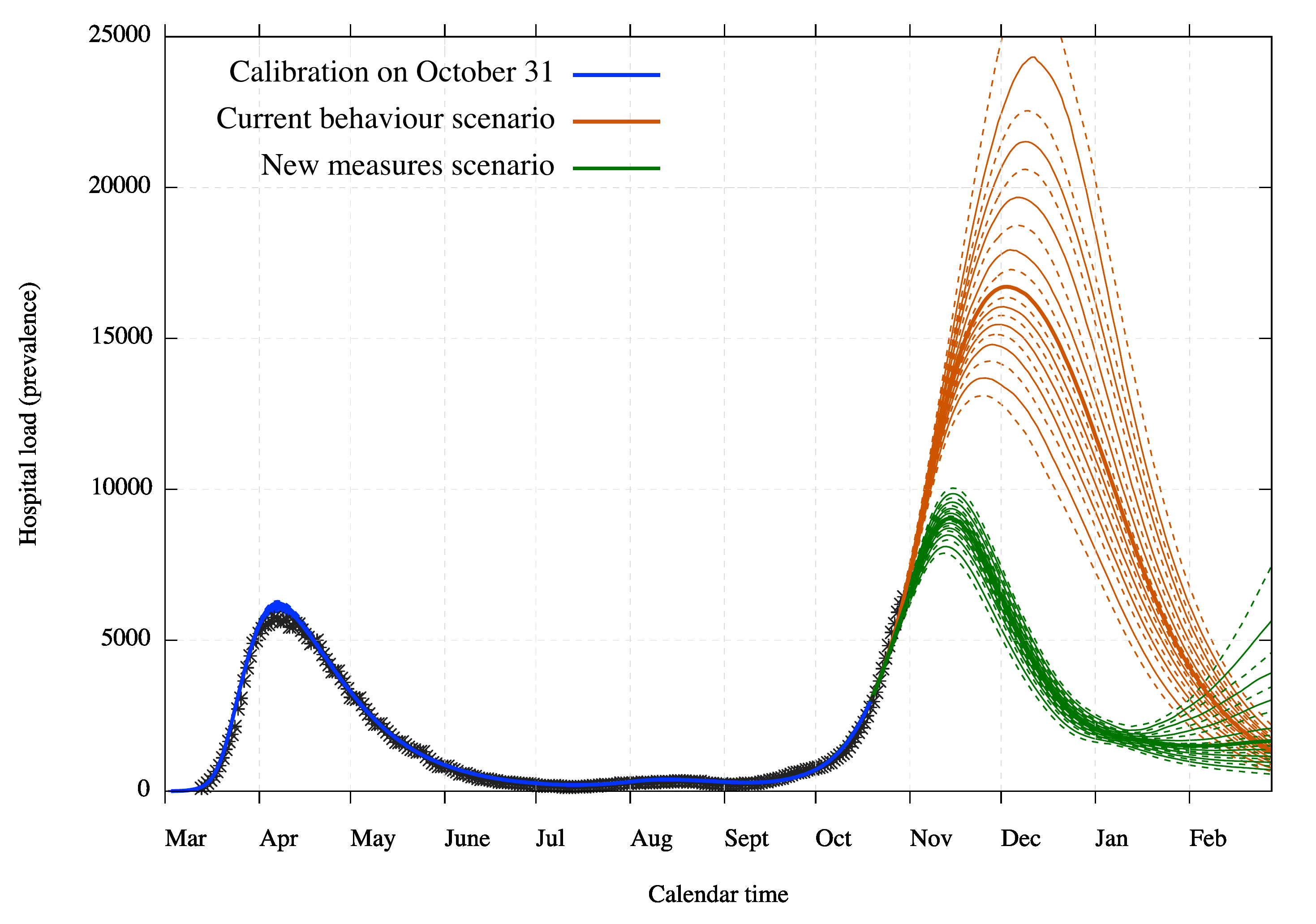}
\caption{Hospital load for two scenarios (with ventiles)}
\label{mid_load}
\end{subfigure}
\begin{subfigure}{.5\textwidth}
\centering
\includegraphics[width=\textwidth]{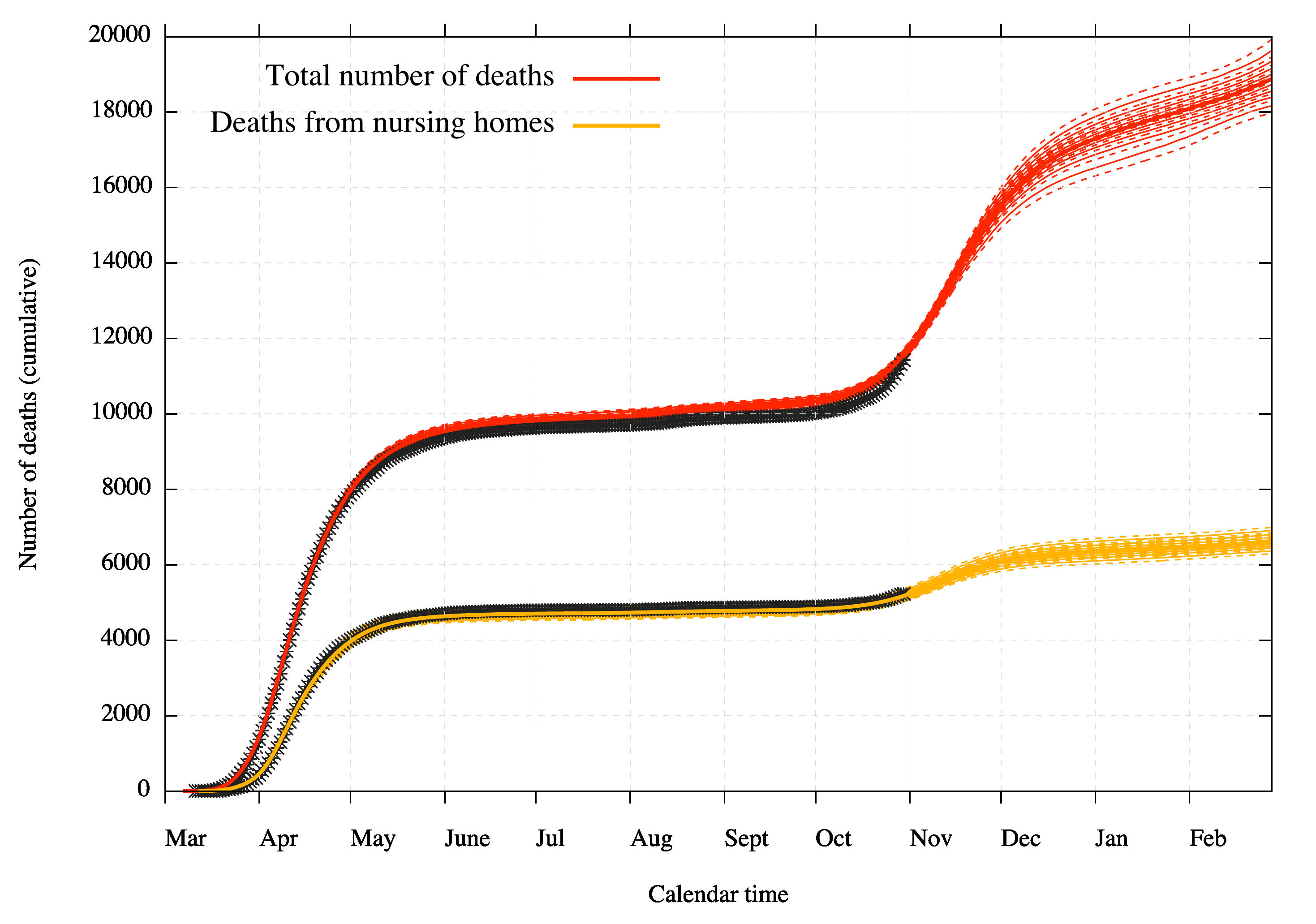}
\caption{Deaths projection for new measures scenario (with ventiles)}
\label{mid_deaths}
\end{subfigure}
\par\bigskip
\par\bigskip
\begin{subfigure}{.5\textwidth}
\centering
\includegraphics[width=\textwidth]{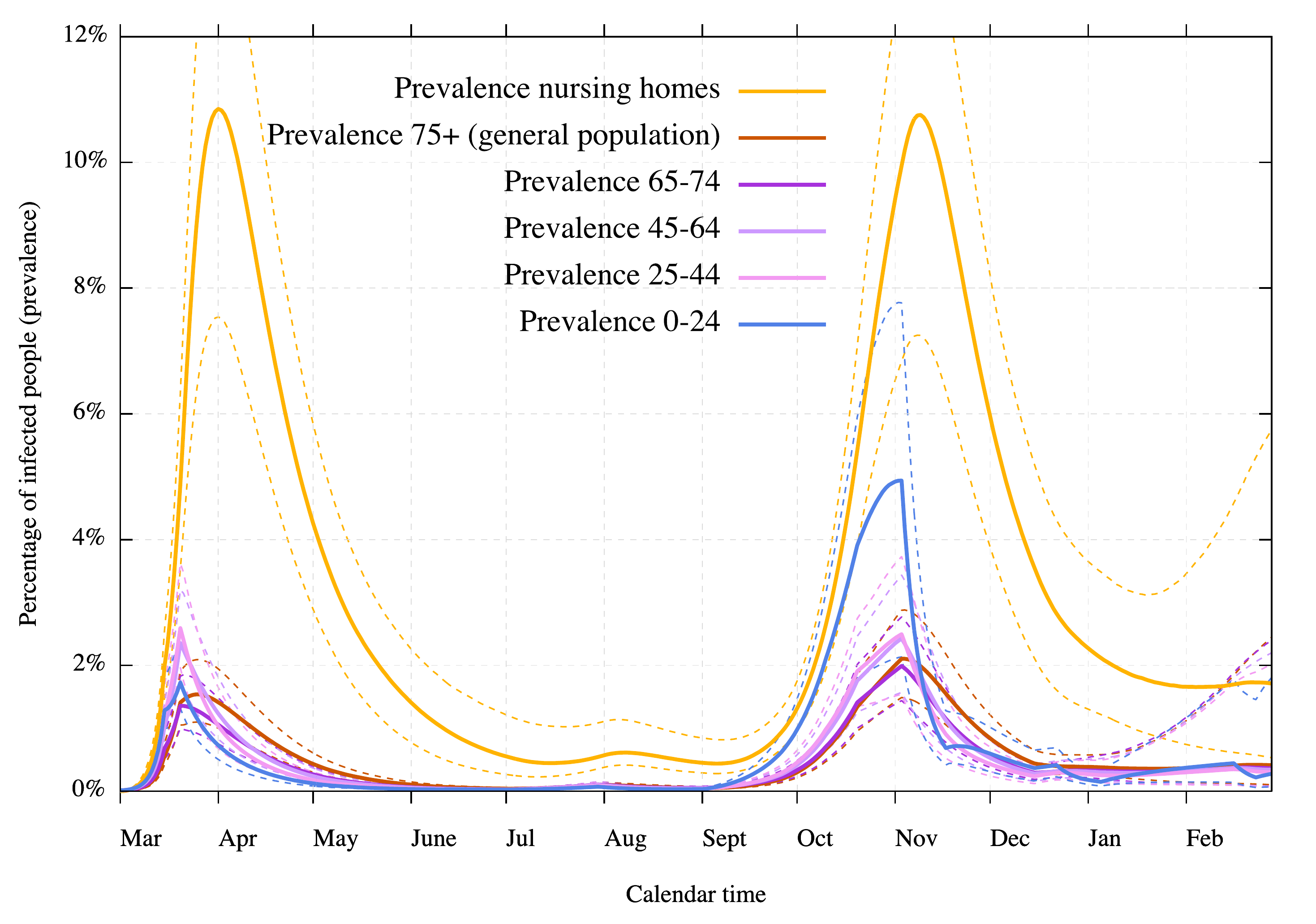}
\caption{Estimated prevalence  for new measures scenario\\ (confidence interval 90\%)}
\label{mid_prev}
\end{subfigure}
\begin{subfigure}{.5\textwidth}
\centering
\includegraphics[width=\textwidth]{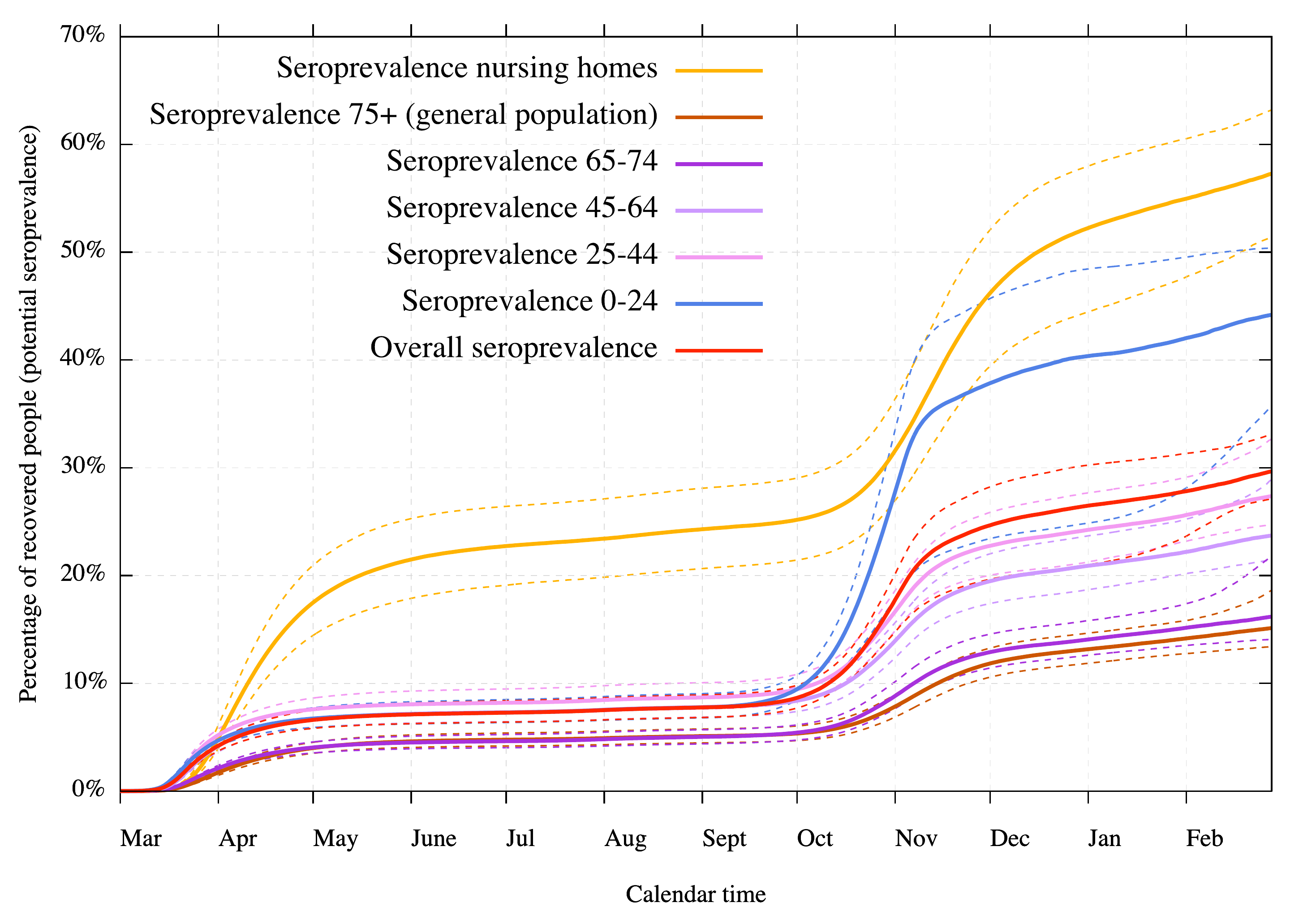}
\caption{Potential seroprevalence (recovered compartments $R_i$) for new measures scenario (confidence interval 90\%)}
\label{mid_sero}
\end{subfigure}
\par\bigskip
\caption{Mid-term scenario with potential effects from new measures applied on October 19 and November 2. The first figure presents a comparison of the hospital load with or without the effects from the new measures. The others figures present the projections on mortality, prevalence and seroprevalence under the new measures. Uncertainty covers both uncertainty about disease parameters and the impact of control measures.}
\label{fig_proj}
\par\bigskip
\end{figure}

In Figure \ref{mid_load}, we present the estimated effect on hospital load from those measures. We note that, according to those measures and to the model, the theoretical maximum capacity of 10000 hospital beds in Belgium should be almost reached but not exceeded. Figure \ref{mid_deaths} presents the expected mortality in case of the new measures scenario. This expected mortality relies on a quality of care that may not be maintained.\\

We must remark that the uncertainty shown in the different projections (Figures \ref{fig_prev}, \ref{fig_proj}, \ref{fig_sc}) is the result of uncertainty on posterior parameter estimates, but those parameters include both disease characteristics and previous reductions on specific contact rates (coefficients $C_*$). For example, the current behaviour scenario in Figure \ref{mid_load} is represented using the uncertainty coming from COVID-19 estimated characteristics as well as the uncertainty coming from the estimation of the current behaviour, while the new measures scenario also uses the uncertainty coming from the estimation of first lockdown contact rates.\\

We can also extrapolate the evolution of the prevalence. In Figure \ref{mid_prev}, we present the estimated percentage of infected people over time for each age class. We can clearly see the effect of mid-March lockdown measures on children and working people. The effect of lockdown measures on older people (especially 75+) is less important since the curve is broken in a less effective manner. Concerning the second wave, we can see that the virus is really present among the very young population due to two complete months of school opening. Prevalence  in this age groups is drastically reduced by the two weeks closure and brought at a lower level than other age classes.\\

In Figure \ref{mid_sero}, we present the estimated percentage of recovered people, hence the estimated percentage of immunity acquired within each age class if we make the assumption that a non-waning immunity is granted to recovered people. Such a lasting immunity is not guaranteed for the moment, but recent studies show that antibodies are present after several months for a large majority of the population \cite{Wajnbergeabd7728}. The seroprevalence is calibrated using blood donor tests results (around 1.3\% on March 30 and 4.7\% on April 14) \cite{SCIENSANO}. Since those tests where only performed on an (almost) asymptomatic population which have not developed COVID-19 symptoms from the past 4 weeks, the model also takes into account immunity coming from the symptomatic population and from nursing homes. Note that we allow a 7 days delay in our model after recovering to be sure of the detectability of the antibodies. Table \ref{table_sero} presents the detail of some seroprevalence estimation.
\begin{table}[htb!]
\footnotesize
\centering
\begin{tabular}{|c|c|c|c|}
\hline
 & global immunity  & among asymptomatic & inside nursing homes\\
 \hline
 March 30 & 2.53\% [2.22\% ; 2.86\%]  & 2.22\% [1.89\% ; 2.52\%]  & 1.42\% [1.12\% ; 1.81\%] \\
 April 14 & 5.16\% [4.(7\% ; 5.76\%]  & 4.24\% [3.66\% ; 4.66\%]  & 8.79\% [7.14\% ; 10.74\%] \\
 October 31 & 16.80\% [14.14\% ; 19.05\%]  & 9.35\% [8.26\% ; 10.10\%]  & 30.71\% [26.20\% ; 35.43\%] \\
 January 1 & 26.53\% [21.02\% ; 30.29\%]  & 17.13 \% [15.21\% ;  18.33\%]  & 52.23\% [44.42\% ; 58.00\%] \\
 \hline
 \end{tabular}
 \caption{Seroprevalence estimations for the new measures scenario}
 \label{table_sero}
 \end{table}

\subsection{Long-term scenarios-based projections}\label{seclong}
The model allows to construct long-term scenarios which are very suitable to study the potential impact from a specific measure. The possibilities are numerous but we present in this section a simple study of the potential impact of an increase in contacts at a specific place (school, family, work and leisure). The increase is perform from January 4, 2021 up to June 30, 2021, when the risk of an emerging third wave is present. We work here with the assumption that there is no modification on the set of susceptible people except from natural infection, hence with the assumption that a non-waning immunity is granted to recovered people. This hypothesis could be modified negatively in the future if the probability of a reinfection is important or positively if the immunity is artificially increased by the arrival of a vaccine. We must emphasise that those scenarios are not real forecasts but only projections under some assumptions. In particular, these projections do not take into account any potential variant of concern with significantly different characteristics.\\

The baseline scenario is the restart of all activities on January 4 with similar transmissions/contacts as in September. Those estimated contacts percentage are $C_\text{school}$=88.2\% [40.5 \%; 99.0\%] for school contacts, $C_\text{home}$=51.4\% [46.9 \%; 54.4\%] for family contacts, $C_\text{work}$=9.3\% [6.0 \%; 14.5\%] for work contacts and $C_\text{leisure}$=31.3\% [21.2 \%; 55.6\%] for leisure contacts. We recall that those percentages do not correspond to the exact number of contacts as determined by the attendance, but to the reduced transmission in comparison to the pre-lockdown period as the result of decrease of contacts but also of sanitary measures. These numbers reflect that transmission is estimated at a very low level at work since sanitary measures and social distancing are more respected than during leisures or among family. The high transmission percentage at school does not necessarily mean that schools are the engine of the virus transmission since most of the student are asymptomatic with a reduced infectiousness, and the uncertainty concerning this parameter is very high.\\

\begin{figure}[!htb]  
\begin{subfigure}{.5\textwidth}
\centering
\includegraphics[width=\textwidth]{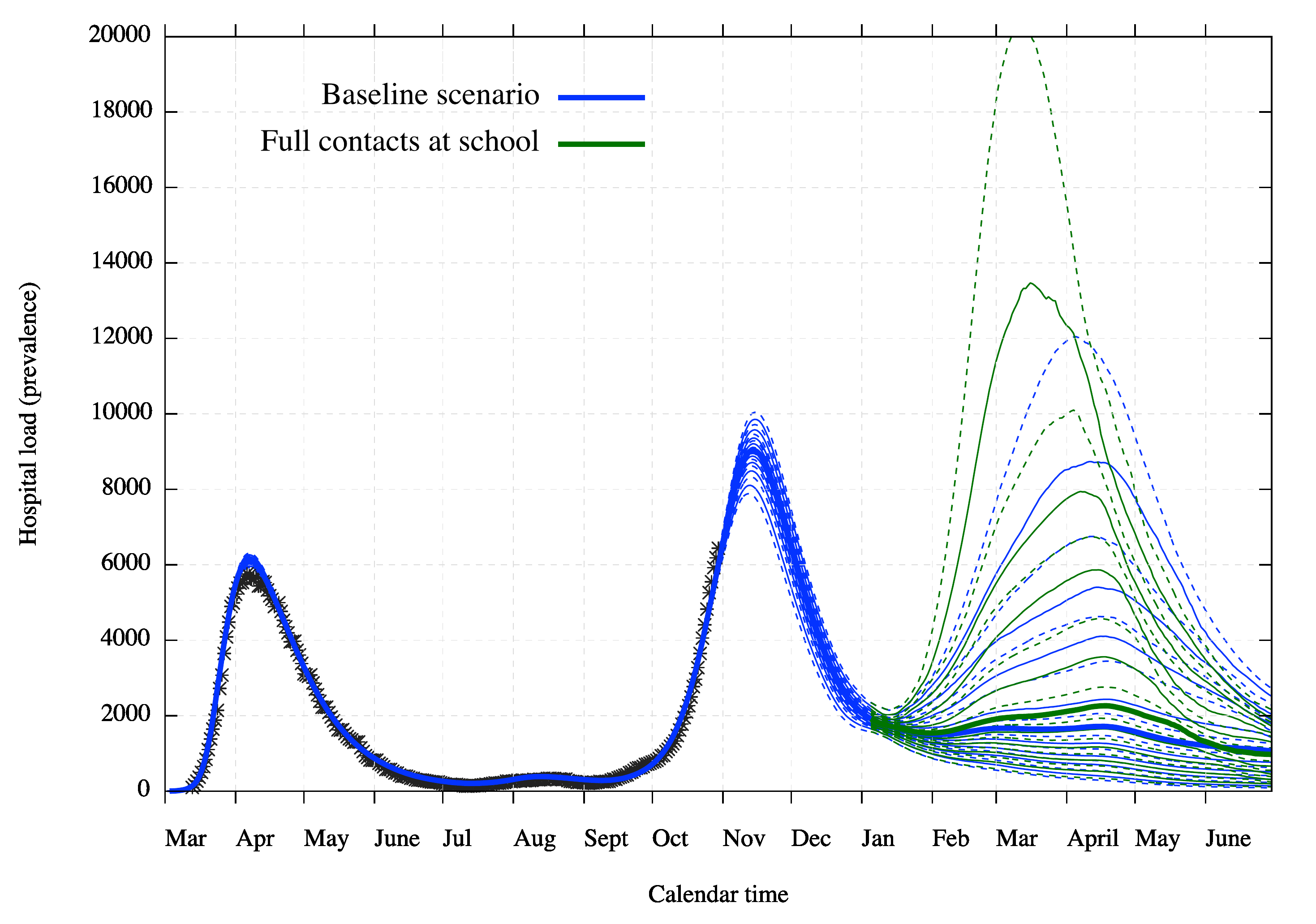}
\caption{Increase in school contacts\\ ($C_\text{school}=100\%$ instead of 88.2\% [40.5 \%; 99.0\%]}
\label{scschool}
\end{subfigure}
\begin{subfigure}{.5\textwidth}
\centering
\includegraphics[width=\textwidth]{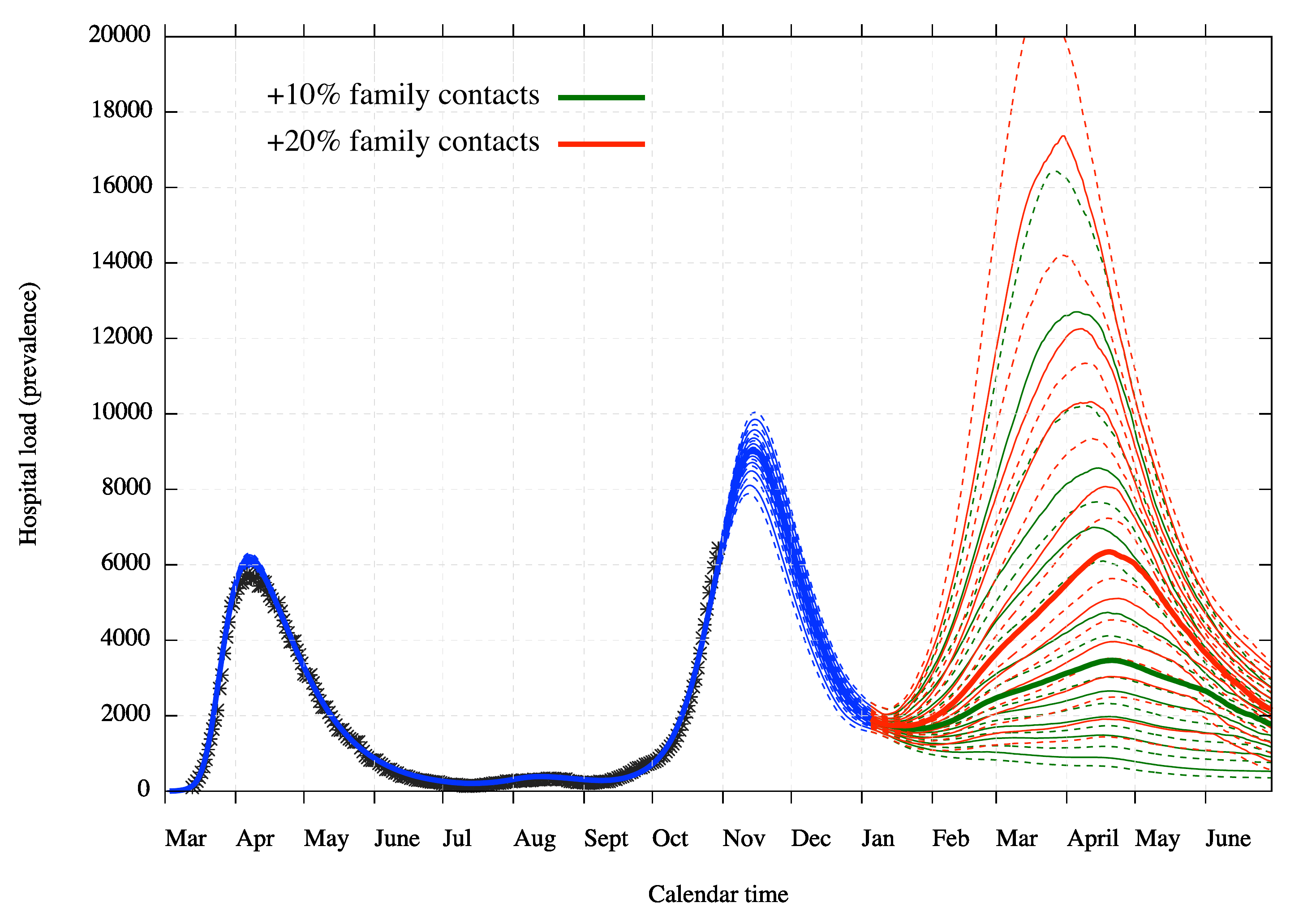}
\caption{Increase in family contacts\\
($C_\text{home}$=51.4\% [46.9 \%; 54.4\%] + 10\% or 20\%)}
\label{schome}
\end{subfigure}
\par\bigskip
\par\bigskip
\begin{subfigure}{.5\textwidth}
\centering
\includegraphics[width=\textwidth]{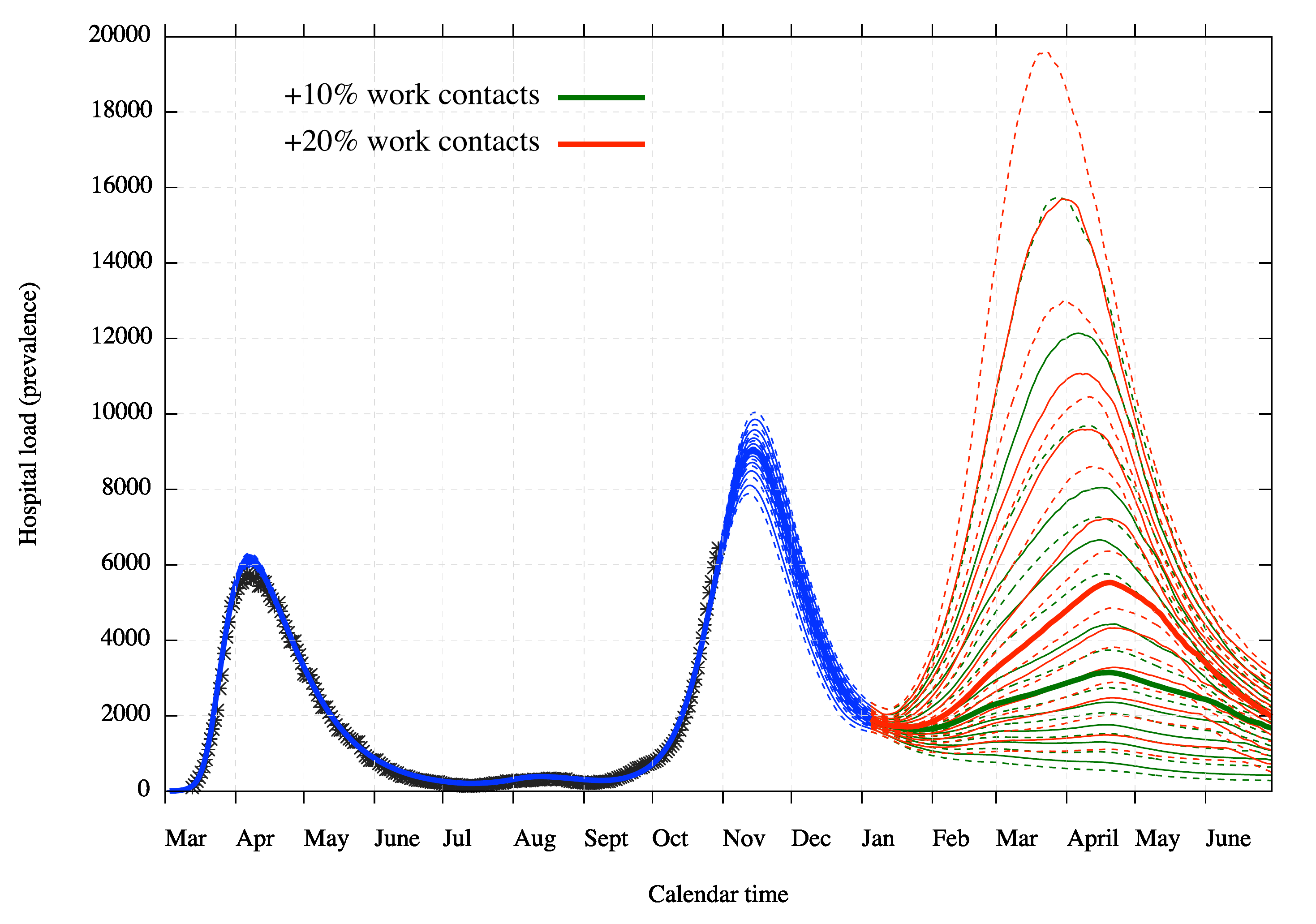}
\caption{Increase in work contacts\\
($C_\text{work}$=9.3\% [6.0 \%; 14.5\%] + 10\% or 20\%)}
\label{scwork}
\end{subfigure}
\begin{subfigure}{.5\textwidth}
\centering
\includegraphics[width=\textwidth]{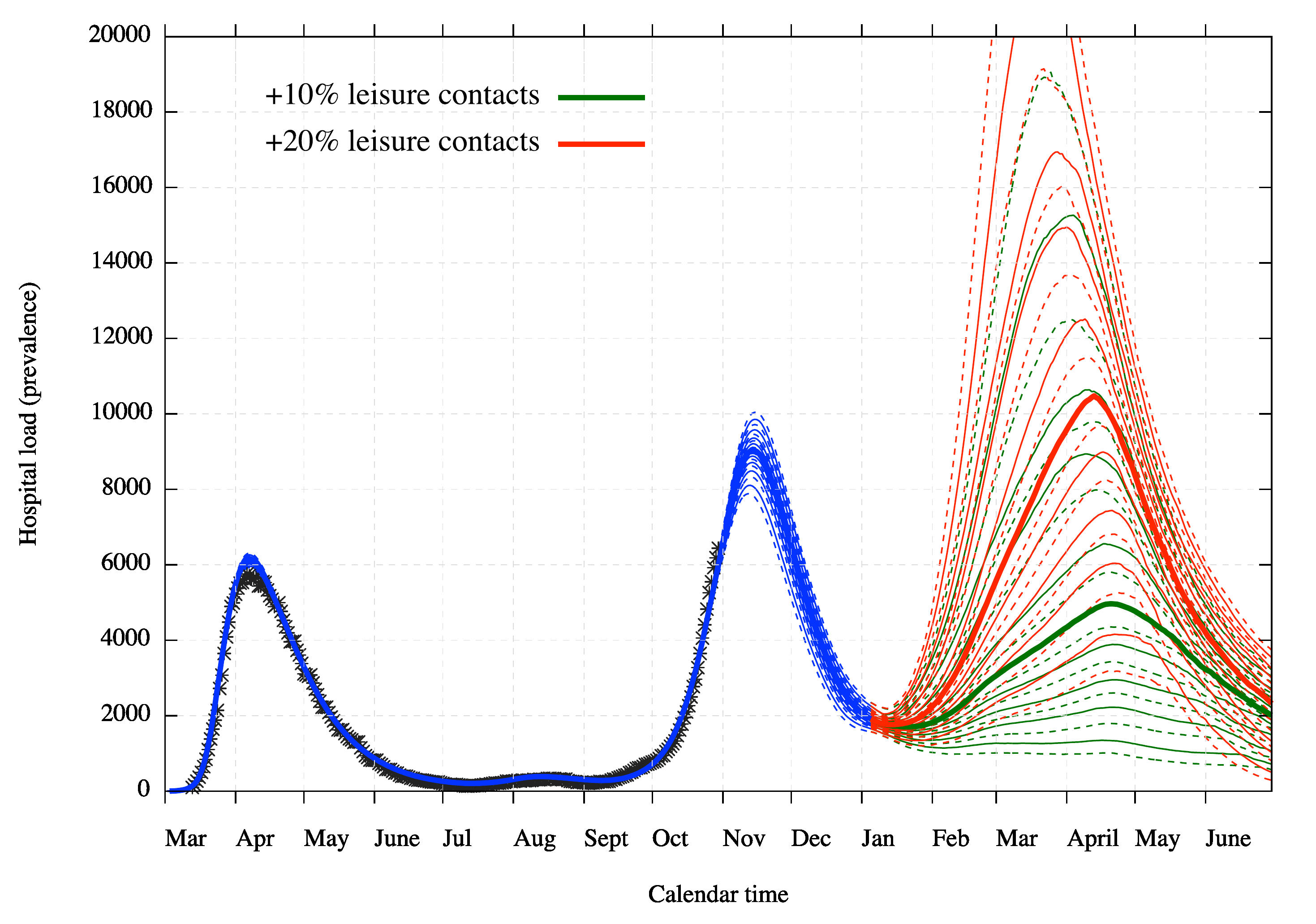}
\caption{Increase in leisure contacts\\
($C_\text{leisure}$=31.3\% [21.2 \%; 55.6\%] + 10\% or 20\%)}
\label{scleisure}
\end{subfigure}
\par\bigskip
\caption{Long-term scenarios with potential isolated effect from an increase in contacts at a specific place (with ventiles). Figure (a) presents the baseline scenario in blue as well as an increase in school contacts in green. The other figures present each time two possibility of increases in contacts in green and red, while the baseline blue scenario was omitted for readability. Uncertainty covers both uncertainty about disease parameters and the impact of control measures.}
\label{fig_sc}
\par\bigskip
\end{figure}

The baseline scenario is presented in Figure \ref{scschool} together with the potential impact of full transmission at school $C_\text{school}=100\%$, hence a transmission without any sanitary measure as well as without any quarantine imposed by the testing and tracing process. We can see that the baseline scenario itself provides a non-zero probability of a third wave but still low. The full contacts at school scenario increases a bit this probability to a reasonable extent.\\

Increases in family contacts, work contacts and leisure contacts are presented in Figures \ref{schome}, \ref{scwork} and \ref{scleisure} with each time a hypothetical increase of $10\%$ or $20\%$ on respectively $C_\text{home}$, $C_\text{work}$ and $C_\text{leisure}$. Those increase must be understood as a non-proportional increase (e.g.~a work increase of $10\%$ corresponds to $C_\text{work}=9.3\%+10\%=19.3\%$). We can clearly see that an increase in leisure contact has the most important effect on the evolution of the epidemic and could lead to a potentially problematic third wave. Full transmission scenarios for family, work or leisure cannot be taken as realistic since they would provide a complete explosion in the absence of vaccine.

\subsection{Conclusion}\label{conclusion}

We have presented an age-structured SEIR-QD type model with a number of improvements compared to others models as a specific consideration for nursing homes, variable parameters and reimportation from travellers. Those improvements were important in order to catch the specificity of the epidemic in Belgium.\\

The model allows to have a good study of the current behaviour of the epidemic, with an estimation of hidden elements like the real prevalence of the virus and the potential evolution of the immunity. More important, the model allows to construct scenarios-based projections in order to estimate the potential impact from new policy measures and can explicitly serve to complement others models for policy makers.\\

However, the model suffers from several limitations which would be important to try to solve in order to better catch the evolution of the epidemic. 
In particular, the model is only capturing an average behaviour resulting in a kind of underestimation of the uncertainty on the real data. This is mainly due to the deterministic nature of the model not capturing daily stochastic realisations. This model captures some key elements as heterogeneity concerning the age structure and the particular role of nursing home but misses other heterogeneous aspects. For example, the lack of spatial consideration is a huge approximation of the reality, even if the Belgian country is small and very connected. Also, the compartmental distinction is limited to asymptomatic and symptomatic while there are several variations of the severity and hospitals are considered as a unique homogeneous element. Furthermore, the lack of refinement inside age classes is a brake on the study of interesting scenarios, as e.g.~studying the separated impact from transmission at primary school, secondary school or university. We must remark however that such a distinction is impossible without sufficiently refined data, and those are not publicly released in Belgium, which is very problematic for quality scientific research.

\section*{Acknowledgement}

The author wants to acknowledge the different members of the Walloon consortium on mathematical model of the COVID-19 epidemic for the numerous discussions, especially Sebastien Clesse, Annick Sartenaer, Alexandre Mauroy, Timoteo Carletti as well as Germain van Bever for statistical discussions. The author wants also to acknowledge the members of the Flemish consortium for the very useful exchanges, models' comparisons and helps on improvement, especially the members of the SIMID-COVID-19 consortium (UHasselt-UAntwerp) and the BIOMATH team (UGent).\\

 This work was supported by the Namur Institute for Complex Systems (naXys) and the Department of Mathematics of the University of Namur, Belgium. The funders had no role in study design, data collection and analysis, decision to publish, or preparation of the manuscript. Computational resources have been provided by the Consortium des \'Equipements de Calcul Intensif (C\'ECI), funded by the Fonds de la Recherche Scientifique de Belgique (F.R.S.-FNRS) under Grant No. 2.5020.11 and by the Walloon Region.

\bibliography{seir-qd}

\begin{thebibliography}{10}
\expandafter\ifx\csname url\endcsname\relax
  \def\url#1{\texttt{#1}}\fi
\expandafter\ifx\csname urlprefix\endcsname\relax\def\urlprefix{URL }\fi
\expandafter\ifx\csname href\endcsname\relax
  \def\href#1#2{#2} \def\path#1{#1}\fi

\bibitem{Rock:2014aa}
K.~Rock, S.~Brand, J.~Moir, M.~J. Keeling, Dynamics of infectious diseases.,
  Rep Prog Phys 77~(2) (2014) 026602.
\newblock \href {https://doi.org/10.1088/0034-4885/77/2/026602}
  {\path{doi:10.1088/0034-4885/77/2/026602}}.

\bibitem{Peng2020.02.16.20023465}
L.~Peng, W.~Yang, D.~Zhang, C.~Zhuge, L.~Hong,
  \href{https://www.medrxiv.org/content/early/2020/02/18/2020.02.16.20023465}{Epidemic
  analysis of covid-19 in china by dynamical modeling}, medRxiv (2020).
\newblock \href {https://doi.org/10.1101/2020.02.16.20023465}
  {\path{doi:10.1101/2020.02.16.20023465}}.
\newline\urlprefix\url{https://www.medrxiv.org/content/early/2020/02/18/2020.02.16.20023465}

\bibitem{Yang2020.03.12.20034595}
W.~Yang, D.~Zhang, L.~Peng, C.~Zhuge, L.~Hong,
  \href{https://www.medrxiv.org/content/early/2020/03/16/2020.03.12.20034595}{Rational
  evaluation of various epidemic models based on the covid-19 data of china},
  medRxiv (2020).
\newblock \href {https://doi.org/10.1101/2020.03.12.20034595}
  {\path{doi:10.1101/2020.03.12.20034595}}.
\newline\urlprefix\url{https://www.medrxiv.org/content/early/2020/03/16/2020.03.12.20034595}

\bibitem{ABRAMS2021100449}
S.~Abrams, J.~Wambua, E.~Santermans, L.~Willem, E.~Kuylen, P.~Coletti,
  P.~Libin, C.~Faes, O.~Petrof, S.~A. Herzog, P.~Beutels, N.~Hens,
  \href{https://www.sciencedirect.com/science/article/pii/S1755436521000116}{Modelling
  the early phase of the belgian covid-19 epidemic using a stochastic
  compartmental model and studying its implied future trajectories}, Epidemics
  35 (2021) 100449.
\newblock \href {https://doi.org/https://doi.org/10.1016/j.epidem.2021.100449}
  {\path{doi:https://doi.org/10.1016/j.epidem.2021.100449}}.
\newline\urlprefix\url{https://www.sciencedirect.com/science/article/pii/S1755436521000116}

\bibitem{Alleman2020.07.17.20156034}
T.~W. Alleman, J.~Vergeynst, E.~Torfs, D.~Illana~Gonzalez, I.~Nopens, J.~M.
  Baetens,
  \href{https://www.medrxiv.org/content/early/2020/07/20/2020.07.17.20156034}{A
  deterministic, age-stratified, extended seird model for assessing the effect
  of non-pharmaceutical interventions on sars-cov-2 spread in belgium}, medRxiv
  (2020).
\newblock \href {https://doi.org/10.1101/2020.07.17.20156034}
  {\path{doi:10.1101/2020.07.17.20156034}}.
\newline\urlprefix\url{https://www.medrxiv.org/content/early/2020/07/20/2020.07.17.20156034}

\bibitem{Willem:2021tl}
L.~Willem, S.~Abrams, P.~J.~K. Libin, P.~Coletti, E.~Kuylen, O.~Petrof,
  S.~M{\o}gelmose, J.~Wambua, S.~A. Herzog, C.~Faes, P.~Beutels, N.~Hens,
  \href{https://doi.org/10.1038/s41467-021-21747-7}{The impact of contact
  tracing and household bubbles on deconfinement strategies for covid-19},
  Nature Communications 12~(1) (2021) 1524.
\newblock \href {https://doi.org/10.1038/s41467-021-21747-7}
  {\path{doi:10.1038/s41467-021-21747-7}}.
\newline\urlprefix\url{https://doi.org/10.1038/s41467-021-21747-7}

\bibitem{Coletti2020.07.20.20157933}
P.~Coletti, P.~Libin, O.~Petrof, L.~Willem, S.~Abrams, S.~A. Herzog, C.~Faes,
  E.~Kuylen, J.~Wambua, P.~Beutels, N.~Hens,
  \href{https://doi.org/10.1186/s12879-021-06092-w}{A data-driven
  metapopulation model for the belgian covid-19 epidemic: assessing the impact
  of lockdown and exit strategies}, BMC Infectious Diseases 21~(1) (2021) 503.
\newblock \href {https://doi.org/10.1186/s12879-021-06092-w}
  {\path{doi:10.1186/s12879-021-06092-w}}.
\newline\urlprefix\url{https://doi.org/10.1186/s12879-021-06092-w}

\bibitem{Willem:2020aa}
L.~Willem, T.~Van~Hoang, S.~Funk, P.~Coletti, P.~Beutels, N.~Hens,
  \href{https://doi.org/10.1186/s13104-020-05136-9}{Socrates: an online tool
  leveraging a social contact data sharing initiative to assess mitigation
  strategies for covid-19}, BMC Research Notes 13~(1) (2020) 293.
\newblock \href {https://doi.org/10.1186/s13104-020-05136-9}
  {\path{doi:10.1186/s13104-020-05136-9}}.
\newline\urlprefix\url{https://doi.org/10.1186/s13104-020-05136-9}

\bibitem{MRMRS}
\href{http://assistance-retraite.be/les-maisons-de-repos-belges-en-quelques-chiffres}{Number
  of nursing homes in belgium} [online].
\newblock webarchive:
  \url{https://web.archive.org/web/20180813023236/http://assistance-retraite.be/les-maisons-de-repos-belges-en-quelques-chiffres}
  [cited 2020].

\bibitem{SCIENSANO}
\href{https://epistat.wiv-isp.be/covid/}{Sciensano: Datasets and
  epidemiological reports} [online] (2020).

\bibitem{NH}
G.~Sophie, J.~L. Belche, J.-f. Moreau, Covid-19 epidemic in the nursing homes
  in belgium, The Journal of Nursing Home Research Science (JNHRS) 6 (2020)
  40--42.
\newblock \href {https://doi.org/10.14283/jnhrs.2020.10}
  {\path{doi:10.14283/jnhrs.2020.10}}.

\bibitem{underreport}
Sciensano,
  \href{https://covid-19.sciensano.be/sites/default/files/Covid19/HOSPITALISATIES%20COVID-19_%20Update%20van%20de%20gegevens_11%20februari%202021.pdf}{Hospitalisations
  et d{\'e}c{\`e}s covid-19 mise {\`a} jour des donn{\'e}es 11 f{\'e}vrier et
  16 mars 2021}.
\newline\urlprefix\url{https://covid-19.sciensano.be/sites/default/files/Covid19/HOSPITALISATIES%20COVID-19_%20Update%20van%20de%20gegevens_11%20februari%202021.pdf}

\bibitem{sciensanohosp}
R.~de~Pauw, B.~Serrien, N.~van Goethem, K.~Blot,
  \href{https://covid-19.sciensano.be/sites/default/files/Covid19/COVID-19_Hospital_epidemiology_Part_1.pdf}{covid-10
  clinical hospital surveillance report}, Tech. rep., Sciensano (2021).
\newline\urlprefix\url{https://covid-19.sciensano.be/sites/default/files/Covid19/COVID-19_Hospital_epidemiology_Part_1.pdf}

\bibitem{KCE}
C.~van~de Voorde, M.~Lev{\`e}bre, P.~Mistiaen, J.~Detonnelaere, L.~Kohn,
  K.~van~den Heede,
  \href{https://kce.fgov.be/sites/default/files/atoms/files/KCE_335_Surge_capacity_during_COVID-19_Belgium_Report_1.pdf}{Assessing
  the management of hospital surge capacity in the first wave of the covid-19
  pandemic in belgium}, Report 335, KCE Belgium Healt Care Knowledge Center
  (2020).
\newline\urlprefix\url{https://kce.fgov.be/sites/default/files/atoms/files/KCE_335_Surge_capacity_during_COVID-19_Belgium_Report_1.pdf}

\bibitem{STATBEL}
\href{https://statbel.fgov.be/en/themes/population/structure-population}{Statbel:
  Structure of the population} [online] (2020).

\bibitem{10.1093/aje/kwj317}
J.~Wallinga, P.~Teunis, M.~Kretzschmar,
  \href{https://doi.org/10.1093/aje/kwj317}{{Using Data on Social Contacts to
  Estimate Age-specific Transmission Parameters for Respiratory-spread
  Infectious Agents}}, American Journal of Epidemiology 164~(10) (2006)
  936--944.
\newblock \href {https://doi.org/10.1093/aje/kwj317}
  {\path{doi:10.1093/aje/kwj317}}.
\newline\urlprefix\url{https://doi.org/10.1093/aje/kwj317}

\bibitem{10.1371/journal.pone.0048695}
L.~Willem, K.~Van~Kerckhove, D.~L. Chao, N.~Hens, P.~Beutels,
  \href{https://doi.org/10.1371/journal.pone.0048695}{A nice day for an
  infection? weather conditions and social contact patterns relevant to
  influenza transmission}, PLOS ONE 7~(11) (2012) 1--7.
\newblock \href {https://doi.org/10.1371/journal.pone.0048695}
  {\path{doi:10.1371/journal.pone.0048695}}.
\newline\urlprefix\url{https://doi.org/10.1371/journal.pone.0048695}

\bibitem{Diekmann:1990aa}
O.~Diekmann, J.~A.~P. Heesterbeek, J.~A.~J. Metz,
  \href{https://doi.org/10.1007/BF00178324}{On the definition and the
  computation of the basic reproduction ratio r0 in models for infectious
  diseases in heterogeneous populations}, Journal of Mathematical Biology
  28~(4) (1990) 365--382.
\newblock \href {https://doi.org/10.1007/BF00178324}
  {\path{doi:10.1007/BF00178324}}.
\newline\urlprefix\url{https://doi.org/10.1007/BF00178324}

\bibitem{doi:10.1098/rsif.2009.0386}
O.~Diekmann, J.~A.~P. Heesterbeek, M.~G. Roberts,
  \href{https://royalsocietypublishing.org/doi/abs/10.1098/rsif.2009.0386}{The
  construction of next-generation matrices for compartmental epidemic models},
  Journal of The Royal Society Interface 7~(47) (2010) 873--885.
\newblock \href {https://doi.org/10.1098/rsif.2009.0386}
  {\path{doi:10.1098/rsif.2009.0386}}.
\newline\urlprefix\url{https://royalsocietypublishing.org/doi/abs/10.1098/rsif.2009.0386}

\bibitem{healthybelgium}
\href{https://www.healthybelgium.be/metadata/hspa/eld4.pdf}{Healthy belgium:
  Performance of the belgian health system -- report 2019} [online] (2019).

\bibitem{ABTO}
\href{https://www.abto.be/wp-content/uploads/2019/10/GfK-Summary-Page-September-2019.pdf}{Abto:
  Association of belgian travel organisers, september 2019 travel trends}
  [online] (2019).

\bibitem{ECDC}
\href{https://www.ecdc.europa.eu/en/publications-data/download-todays-data-geographic-distribution-covid-19-cases-worldwide}{Ecdc:
  European centre for disease prevention and control, daily number of new
  reported cases of covid-19 by country worldwide} [online] (2020).

\bibitem{metropolis}
N.~Metropolis, A.~W. Rosenbluth, M.~N. Rosenbluth, A.~H. Teller, E.~Teller,
  \href{https://doi.org/10.1063/1.1699114}{Equation of state calculations by
  fast computing machines}, The Journal of Chemical Physics 21~(6) (1953)
  1087--1092.
\newblock \href {http://arxiv.org/abs/https://doi.org/10.1063/1.1699114}
  {\path{arXiv:https://doi.org/10.1063/1.1699114}}, \href
  {https://doi.org/10.1063/1.1699114} {\path{doi:10.1063/1.1699114}}.
\newline\urlprefix\url{https://doi.org/10.1063/1.1699114}

\bibitem{hilbe_2014}
J.~M. Hilbe, Modeling Count Data, Cambridge University Press, 2014.
\newblock \href {https://doi.org/10.1017/CBO9781139236065}
  {\path{doi:10.1017/CBO9781139236065}}.

\bibitem{Lesaffre}
E.~Lesaffre, A.~B. Lawson, Bayesian biostatistics, Wiley, 2012.

\bibitem{githubcorona}
{Franco Nicolas},
  \href{https://github.com/nicolas-franco-unamur/corona_seiirqd}{corona seiirqd
  github repository}.
\newline\urlprefix\url{https://github.com/nicolas-franco-unamur/corona_seiirqd}

\bibitem{Wajnbergeabd7728}
A.~Wajnberg, F.~Amanat, A.~Firpo, D.~R. Altman, M.~J. Bailey, M.~Mansour,
  M.~McMahon, P.~Meade, D.~R. Mendu, K.~Muellers, D.~Stadlbauer, K.~Stone,
  S.~Strohmeier, V.~Simon, J.~Aberg, D.~L. Reich, F.~Krammer, C.~Cordon-Cardo,
  \href{https://science.sciencemag.org/content/early/2020/10/27/science.abd7728}{Robust
  neutralizing antibodies to sars-cov-2 infection persist for months}, Science
  (2020).
\newblock \href {https://doi.org/10.1126/science.abd7728}
  {\path{doi:10.1126/science.abd7728}}.
\newline\urlprefix\url{https://science.sciencemag.org/content/early/2020/10/27/science.abd7728}

\end{thebibliography}

\begin{appendix}

\clearpage
\section{Supplementary material: model details, timeline and estimated parameters}\label{secappend}

In this Appendix, we provide some technical details concerning the construction of the model. 

\subsection{Terminology and parameters description}

A description of the terminology used for the compartmental model is presented in Table \ref{compartments} and a description of the model parameters in Table \ref{table_par}. Parameters without age class index $i$ are assumed similar for all classes while those with age class index $i$ are class-dependent.

\begin{table}[ht!]
\centering
{\scriptsize
\begin{tabular}{|c|c|c|}
\hline
$S_i$ &Susceptible&People who have never been infected and are a priori susceptible to be infected\\
\hline
$E_i$ & Exposed & \makecell{People who have just been infected but are without any symptom\\ and still not infectious (latent period)}\\
\hline
$I^A_i$ & Asymptomatic Infectious & \makecell{This is the part of the exposed people who fall into a continuously asymptomatic disease,\\ which are infectious but with a reduced infectiousness due to their\\ asymptomatic status and directly fall into the recovered status after that period}\\
\hline
$I^P_i$ & Presymptomatic Infectious & \makecell{This is the other part of the exposed people who fall into a symptomatic disease,\\ but symptoms do not appear directly,\\ hence there is an intermediate stage where people become infectious\\ but still without any symptom and with a still reduced infectinousness}\\
\hline
$I^S_i$ & Symptomatic Infectious & \makecell{Real disease period where the infectinousness is higher\\ People in this compartment will eventually fall either in a recovered status\\ or will be hospitalised, and concerning nursing home,\\ a significant part of them will die without hospitalisation}\\
\hline
$Q_i$ & Hospitalised & \makecell{Hospitalised people are considered as in quarantine for the model,\\ since their contacts are almost inexistant}\\
\hline
$D_i$ & Deceased & \makecell{Deaths from the general population are assumed only coming from hospitalised people\\
There is a small $1\%$ of exceptions which is not taken into consideration here\\
However, deaths from nursing homes are taken into consideration\\ and separated from deaths coming from hospitals}\\
\hline
$R_i$ & Recovered & \makecell{People who recovered from the disease, from asymptomatic ones,\\ symptomatic ones or from the hospital, and are assumed here immune for the future}\\
\hline
\end{tabular}}
\caption{Description of the 8 compartments of the model according to the different possible stages of the disease.}
\label{compartments}
\end{table}

\begin{table}[ht!]
\centering
\scriptsize
\begin{tabular}{|c|c|c|l|}
\hline
Parameter & \# & Unit & Description \\
\hline
$p_0$ &1&--& proportion of infected people on day 1 (initial condition)\\
$\lambda_a$ &1&--& transmission probability from asymptomatic or presymptomatic infectious people\\
$\lambda_s$ &1&--& transmission probability from symptomatic infectious people\\
$\sigma$ &1&day$^{-1}$& rate at which an exposed person becomes contagious (inverse of latent period duration)\\
$\tau$ &1&day$^{-1}$& rate at which a presymptomatic person becomes symptomatic\\
 ${p_a}_i$ &6&--& probability of a completely asymptomatic disease\\
 $\delta_i$ &6&day$^{-1}$& rate at which a symptomatic person develops heavy symptoms and is hospitalised\\
 ${\gamma_a}_i$ &6&day$^{-1}$& rate at which a person recovers from asymptomatic disease\\
 ${\gamma_s}_i$ &6&day$^{-1}$& rate at which a person recovers from symptomatic disease\\
 ${\gamma_q}_i(t)$ &6&day$^{-1}$& variable rate at which a person recovers from hospital (using the "recovery" logistic function)\\
 ${\gamma_q}_i$ &6&day$^{-1}$& baseline rate at which a person recovers from hospital (coefficient of the "recovery" logistic function)\\
 $P_\text{recovery}$ &1&--& percentage of maximal improvement of the "recovery" logistic function\\
 $\mu_\text{recovery}$ &1&days& midpoint of the "recovery" logistic function\\
 $s_\text{recovery}$  &1&days& steepness$^{-1}$ of the "recovery" logistic function\\
 $r_i(t)$ &6&day$^{-1}$& variable rate at which a person dies from hospital (using the "recovery" logistic function)\\
 $r_i$ &6&day$^{-1}$& baseline rate at which a person dies from hospital (coefficient of the "recovery" logistic function)\\
 $\tilde r_h(t)$ &6&day$^{-1}$& variable rate at which a person dies directly from nursing home (using the "hosp" logistic function)\\
 $\tilde r_h$ &6&day$^{-1}$& baseline rate at which a person dies directly from nursing home (coefficient of the "hosp" logistic function)\\
  $\mu_\text{hosp}$ &1&people& midpoint of the "hosp" logistic function\\
 $s_\text{hosp}$  &1&people& steepness$^{-1}$ of the "hosp" logistic function\\
 $\text{delay}$  &1&days& time shift of the "hosp" logistic function\\
 SUPP$_{hosp}$  &1&--& percentage of underreporting in new hospitalisations incidence\\
$P_{cor} $  &1&--& percentage of COVID-19 related deaths among reported deaths from nursing homes \\
 $P_{th}$   &1&--&  coefficient of the probability of nursing home infection from general population before lockdown\\
$P^\prime_{th}$  &1&--&   coefficient of the probability of nursing home infection from general population during and after lockdown\\
 $M_{ij} $ &25&day$^{-1}$& social contact matrix (contact rate of individuals of class $i$ from an individual of class $j$) \\
 $M_{ij*} $ &100&day$^{-1}$& social contact matrices per specific location (home, work, school, leisure) \\
$m_h$  &1&day$^{-1}$& contact rate inside nursing homes\\
$C_\text{reimp}$  &1& --& global coefficient for the estimation of infected travellers \\
$C_\text{*}$  &4+&--&  transmission reduction for a specific location (home, work, school, leisure) during a specific period \\
\hline
\end{tabular}
\caption{Description of the parameters of the model}
\label{table_par}
\end{table}

\newpage
\subsection{Timeline and intervention measures}

According to the start of the pandemic in March 2020, the Belgian government began to apply several non pharmaceutical interventions (NPI) in order to reduce the transmission of the virus and to protect health care capacities. In this subsection, we detail how those interventions have been taken into consideration within our model. Those NPI are modelled by the four generic coefficients of the contact matrices: $C_\text{home}$ for transmission within the family (household and nearby family), $C_\text{work}$ for transmission during work and travels, $C_\text{school}$ for transmission at school and $C_\text{leisure}$ for transmission during leisures or other activities. All those coefficients are considered at maximal value 1 during the pre-lockdown period. Then there are assumed to be at 0 if the sector is completely closed (which only happens which school) or estimated by the model according to specific parameters (whose estimated values are detailed in Table \ref{list_par}).\\

The lockdown during the first wave took place in two steps. From March 14, the government imposed the closure of schools, shops and of all leisures activities, which is modelled by $C_\text{school}=0$ and $C_\text{leisure}=C_\text{leisurelock}$, an estimated specific parameter catching the reduction of the transmission for leisures and others activities. Then from March 18 midday (assumed March 19 here), additional measures were taken imposing stricter measures of physical distancing, with teleworking mandatory whenever possible and all travels restricted to specific essential tasks. From this period, the new transmission at work is modelled by $C_\text{work}=C_\text{worklock}$ and between family members by $C_\text{home}=C_\text{homelock}$ to catch the reduction of contacts with non-household members.\\

The lockdown release was planned with several phases 1A-B, 2, 3 and 4 during May to July 2020 (cf.~Table \ref{list_time} for specific dates). Concerning family and closed contacts, the concept of social bubble \cite{Willem:2021tl} has been implemented, with initially only few contacts (bubble of additional 2 people) and a bubble of maximum 10 people on phase 3. This is estimated by $C_\text{home}=C_\text{homeunlock}$ on phase 3 with an intermediate state $C_\text{home}=(C_\text{homelock}+C_\text{homeunlock})/2$ on previous phases. A further extension of the bubble on phase 4 is modelled by $C_\text{home}=2 C_\text{homeunlock}-C_\text{homelock}$. Concerning works, a progressive reopening took place on phase 1A-B (industries and professional services on 1A and all shops on 1B) modelled by the estimation $C_\text{work}=C_\text{workunlock}$ on phase 1B and the intermediate point on phase 1A. Schools have also a progressive partial opening on phase 2 and 3, modelled by $C_\text{school}=C_\text{schoolunlock}$ on the maximal point on phase 3 and a progressive 20\%-40\%-60\% before. Leisures remained closed until phase 3 in June with limited people and activities modelled by $C_\text{leisure}=C_\text{leisurejune}$ and in phase 4 in July with more people and activities modelled by $C_\text{leisure}=C_\text{leisurejuly}$. \\

Due to a restarting epidemic at the end of July, the government decided to backtrack on some measures, restricting again closed and outside contacts. This is implemented by a backtrack to $C_\text{home}=C_\text{homeunlock}$ and a new estimation of non-work related activities $C_\text{leisure}=C_\text{leisureaug}$. Restarting activities in September are modelled by $C_\text{leisure}=C_\text{leisuresept}$ (since more leisure activities occur during non-holidays periods) and by an estimation of the full opening of schools $C_\text{school}=C_\text{schoolsept}$. The social contact bubble is once more relaxed $C_\text{home}=2 C_\text{homeunlock}-C_\text{homelock}$ with a backtrack end of September $C_\text{home}=C_\text{homeunlock}$.\\

A summary of the Belgian policy timeline with all corresponding social contact matrices coefficients is presented in Table \ref{list_time}. The Table \ref{list_time_sce} contains the social contact matrices coefficients used for the second lockdown scenario described in Subsection \ref{secmid}.

\begin{table}[ht!]
\centering{\tiny
\setlength{\tabcolsep}{1pt}
\begin{tabular}{|c|c|c|c|c|c|}
\hline
Timeline & Summary  &$C_\text{home}$ & $C_\text{work}$& $C_\text{school}$ & $C_\text{leisure}$\\[2px]
\hline
Pre-lockdown: March 1 $\rightarrow$  13 & everything is open & 1 & 1 & 1 & 1 \\[2px]
Half-lockdown: March 14 $\rightarrow$ 18 & schools and all leisures closed & 1 & 1 & 0 & $C_\text{leisurelock}$ \\[2px]
Lockdown: March 19 $\rightarrow$ May 3 & teleworking + travel restrictions  & $C_\text{homelock}$ & $C_\text{worklock}$ & 0 & $C_\text{leisurelock}$ \\[2px]
Phase 1A: May 4 $\rightarrow$  10 & shops partially reopen + few contacts  & $\frac{C_\text{homelock}+C_\text{homeunlock}}{2}$ & $\frac{C_\text{worklock}+C_\text{workunlock}}{2}$ & 0 & $C_\text{leisurelock}$ \\[2px]
Phase 1B: May 11 $\rightarrow$  17 & all shops and compagnies reopen  & $\frac{C_\text{homelock}+C_\text{homeunlock}}{2}$ & $C_\text{workunlock}$ & 0 & $C_\text{leisurelock}$ \\[2px]
Phase 2: May 18 $\rightarrow$  24 & progressive partial opening of schools & $\frac{C_\text{homelock}+C_\text{homeunlock}}{2}$ & $C_\text{workunlock}$ & $0.2  C_\text{schoolunlock}$ & $C_\text{leisurelock}$ \\[2px]
Phase 2: May 25 $\rightarrow$ June 1 & progressive partial opening of schools & $\frac{C_\text{homelock}+C_\text{homeunlock}}{2}$& $C_\text{workunlock}$ & $0.4  C_\text{schoolunlock}$ & $C_\text{leisurelock}$ \\[2px]
Phase 2: June 2 $\rightarrow$  7  & progressive partial opening of schools & $\frac{C_\text{homelock}+C_\text{homeunlock}}{2}$ & $C_\text{workunlock}$ & $0.6  C_\text{schoolunlock}$ & $C_\text{leisurelock}$ \\[2px]
Phase 3: June 8 $\rightarrow$  30  & Schools partially opened + leisures  & $C_\text{homeunlock}$ & $C_\text{workunlock}$ & $C_\text{schoolunlock}$ & $C_\text{leisurejune}$ \\[2px]
Phase 4: July 1 $\rightarrow$  28  & Cultural events  + Social contacts   & $2 C_\text{homeunlock}-C_\text{homelock}$ & $C_\text{workunlock}$ & 0 & $C_\text{leisurejuly}$ \\[2px]
Backtrack: July 29 $\rightarrow$ August 31  & Social contacts restricted  & $C_\text{homeunlock}$ & $C_\text{workunlock}$ & 0 & $C_\text{leisureaug}$ \\[2px]
September 1 $\rightarrow$  24  & bubble optional & $2 C_\text{homeunlock}-C_\text{homelock}$ & $C_\text{workunlock}$ & $C_\text{schoolsept} $ & $ C_\text{leisuresept} $ \\[2px]
September 25 $\rightarrow$ October 18  & bubble mandatory& $C_\text{homeunlock}$ & $C_\text{workunlock}$ &$C_\text{schoolsept} $ & $ C_\text{leisuresept} $ \\[2px]
\hline
\end{tabular}	
}
\caption{Belgian policy timeline and corresponding social contact matrices coefficients}
\label{list_time}
\end{table}

\begin{table}[ht!]
\centering{\tiny
\setlength{\tabcolsep}{1pt}
\begin{tabular}{|c|c|c|c|c|}
\hline
Timeline  &$C_\text{home}$ & $C_\text{work}$& $C_\text{school}$ & $C_\text{leisure}$\\[2px]
\hline
October 19 $\rightarrow$ November 1  &  $\frac{C_\text{homelock}+C_\text{homeunlock}}{2}$ & $\frac{C_\text{worklock}+C_\text{workunlock}}{2}$ &$C_\text{schoolsept} $ & $\frac{C_\text{leisurelock}+C_\text{leisuresept}}{2}$ \\[2px]
November 2 $\rightarrow$ November 15  & $C_\text{homelock}$ & $C_\text{worklock}$ &$0 $ & $ C_\text{leisurelock} $ \\[2px]
November 16 $\rightarrow$ December 13  & $C_\text{homelock}$ & $C_\text{worklock}$ &$5/6\,C_\text{schoolsept} $ & $ C_\text{leisurelock} $ \\[2px]
December 14 $\rightarrow$ December 20  & $C_\text{homeunlock}$ & $C_\text{workunlock}$ &$5/6\,C_\text{schoolsept} $ & $ C_\text{leisuresept} $ \\[2px]
December 21 $\rightarrow$ January 3  & $C_\text{homeunlock}$ & $C_\text{workunlock}$ &$0 $ & $ C_\text{leisuresept} $ \\[2px]
January 4 $\rightarrow$ $\dots$  & $C_\text{homeunlock}$ & $C_\text{workunlock}$ &$5/6\,C_\text{schoolsept} $ & $ C_\text{leisuresept} $ \\[2px]
\hline
\end{tabular}	
}
\caption{Social contact matrices coefficients used for second lockdown scenario}
\label{list_time_sce}
\end{table}

\newpage\vspace*{0.5cm}
\subsection{Estimated parameters}
Table \ref{list_reimp} shows the estimated number of reimportations of COVID-19 per day during the holidays period.  The complete list of estimated parameters from the calibration on October 31, 2020 data is given in Table \ref{list_par}.

\begin{table}[ht!]
\centering
{\scriptsize
\begin{tabular}{|cc|cc|cc|}
\hline
Date & People infected & Date & People infected & Date & People infected \\
\hline
07/01/20	&	44.7	[25.4	;	60.2]	&	08/01/20	&	91.5	[52.0	;	123.3]	&	09/01/20	&	197.2	[112.0	;	265.5]	\\
07/02/20	&	45.5	[25.9	;	61.3]	&	08/02/20	&	92.4	[52.5	;	124.5]	&	09/02/20	&	202.8	[115.2	;	273.1]	\\
07/03/20	&	46.1	[26.2	;	62.1]	&	08/03/20	&	99.5	[56.5	;	134.1]	&	09/03/20	&	209.7	[119.2	;	282.5]	\\
07/04/20	&	44.9	[25.5	;	60.5]	&	08/04/20	&	110.5	[62.8	;	148.9]	&	09/04/20	&	215.1	[122.2	;	289.7]	\\
07/05/20	&	41.9	[23.8	;	56.4]	&	08/05/20	&	115.5	[65.6	;	155.5]	&	09/05/20	&	221.6	[125.9	;	298.5]	\\
07/06/20	&	43.2	[24.5	;	58.1]	&	08/06/20	&	120.5	[68.5	;	162.3]	&	09/06/20	&	228.7	[130.0	;	308.1]	\\
07/07/20	&	46.6	[26.5	;	62.8]	&	08/07/20	&	126.9	[72.1	;	171.0]	&	09/07/20	&	238.8	[135.7	;	321.7]	\\
07/08/20	&	46.5	[26.4	;	62.7]	&	08/08/20	&	131.0	[74.4	;	176.5]	&	09/08/20	&	243.9	[138.6	;	328.5]	\\
07/09/20	&	48.1	[27.3	;	64.8]	&	08/09/20	&	132.1	[75.1	;	177.9]	&	09/09/20	&	250.7	[142.4	;	337.7]	\\
07/10/20	&	50.9	[28.9	;	68.5]	&	08/10/20	&	137.1	[77.9	;	184.7]	&	09/10/20	&	256.8	[145.9	;	345.8]	\\
07/11/20	&	46.3	[26.3	;	62.3]	&	08/11/20	&	146.9	[83.5	;	197.9]	&	09/11/20	&	264.4	[150.2	;	356.1]	\\
07/12/20	&	45.6	[25.9	;	61.5]	&	08/12/20	&	152.3	[86.5	;	205.1]	&	09/12/20	&	268.4	[152.5	;	361.4]	\\
07/13/20	&	50.1	[28.5	;	67.5]	&	08/13/20	&	164.4	[93.4	;	221.4]	&	09/13/20	&	276.2	[157.0	;	372.1]	\\
07/14/20	&	52.0	[29.5	;	70.0]	&	08/14/20	&	172.3	[97.9	;	232.1]	&	09/14/20	&	283.0	[160.8	;	381.1]	\\
07/15/20	&	51.3	[29.1	;	69.1]	&	08/15/20	&	177.1	[100.6	;	238.5]	&	09/15/20	&	289.4	[164.4	;	389.8]	\\
07/16/20	&	53.6	[30.4	;	72.1]	&	08/16/20	&	186.2	[105.8	;	250.7]	&	09/16/20	&	148.5	[84.4	;	200.1]	\\
07/17/20	&	55.5	[31.5	;	74.7]	&	08/17/20	&	207.2	[117.8	;	279.1]	&	09/17/20	&	151.9	[86.3	;	204.6]	\\
07/18/20	&	56.5	[32.1	;	76.1]	&	08/18/20	&	199.6	[113.4	;	268.9]	&	09/18/20	&	157.0	[89.2	;	211.5]	\\
07/19/20	&	56.8	[32.2	;	76.4]	&	08/19/20	&	208.8	[118.6	;	281.2]	&	09/19/20	&	160.8	[91.4	;	216.6]	\\
07/20/20	&	64.8	[36.8	;	87.3]	&	08/20/20	&	219.5	[124.7	;	295.6]	&	09/20/20	&	165.0	[93.8	;	222.3]	\\
07/21/20	&	70.1	[39.8	;	94.4]	&	08/21/20	&	233.5	[132.7	;	314.5]	&	09/21/20	&	171.0	[97.1	;	230.3]	\\
07/22/20	&	73.2	[41.6	;	98.6]	&	08/22/20	&	239.6	[136.1	;	322.7]	&	09/22/20	&	173.0	[98.3	;	233.0]	\\
07/23/20	&	79.7	[45.3	;	107.4]	&	08/23/20	&	249.2	[141.6	;	335.7]	&	09/23/20	&	177.4	[100.8	;	238.9]	\\
07/24/20	&	85.0	[48.3	;	114.5]	&	08/24/20	&	279.6	[158.9	;	376.6]	&	09/24/20	&	181.3	[103.0	;	244.2]	\\
07/25/20	&	87.2	[49.5	;	117.4]	&	08/25/20	&	279.1	[158.6	;	375.9]	&	09/25/20	&	186.8	[106.1	;	251.6]	\\
07/26/20	&	87.5	[49.7	;	117.9]	&	08/26/20	&	290.0	[164.8	;	390.6]	&	09/26/20	&	192.3	[109.2	;	259.0]	\\
07/27/20	&	98.0	[55.7	;	132.0]	&	08/27/20	&	301.3	[171.2	;	405.9]	&	09/27/20	&	196.1	[111.4	;	264.1]	\\
07/28/20	&	104.3	[59.3	;	140.5]	&	08/28/20	&	317.7	[180.5	;	427.9]	&	09/28/20	&	202.2	[114.9	;	272.3]	\\
07/29/20	&	110.2	[62.6	;	148.5]	&	08/29/20	&	329.3	[187.1	;	443.6]	&	09/29/20	&	202.4	[115.0	;	272.6]	\\
07/30/20	&	115.9	[65.8	;	156.1]	&	08/30/20	&	335.0	[190.3	;	451.2]	&	09/30/20	&	203.8	[115.8	;	274.6]	\\
07/31/20	&	123.8	[70.4	;	166.8]	&	08/31/20	&	354.0	[201.1	;	476.8]	\\							
\hline
\end{tabular}
}
\caption{Estimation of reimportations per day of COVID-19 during the holidays period (median and 90\% confidence interval)}
\label{list_reimp}
\end{table}

\begin{table}[ht!]
\centering
{\tiny
\begin{tabular}{|c|c|c|c|c|c|c|}
\hline
Parameter& Short description & Prior (SD)  & Step (SD) & Mean & Median & 90\% confidence interval\\
\hline
$p_0$  	&	 initial value  	&	 $0.0002 \pm \num{2e-5}$  	&	 $\num{2e-7}$  	&	0,000129206698	&	0,000120969179	&	[0.000075026768	;	0.000213341823]
\\ $\lambda_a$ 	&	 transmission (asympt)  	&	 $0.08 \pm \num{5e-3}$  	&	 $\num{5e-5}$  	&	0,058682046841	&	0,058507251651	&	[0.053664079039	;	0.064515921481]
\\ $\lambda_s$ 	&	 transmission (sympt)  	&	  $0.08 \pm \num{5e-3}$  	&	 $\num{5e-5}$  	&	0,066893021074	&	0,066090460437	&	[0.059186574978	;	0.077023561314]
\\ $\sigma$ 	&	 latent period$^{-1}$  	&	 $0.5 \pm \num{5e-2}$  	&	 $\num{5e-4}$  	&	0,72155730196	&	0,737642210086	&	[0.502332749238	;	0.886673410853]
\\ $\tau$ 	&	 presympt period$^{-1}$  	&	 $0.2 \pm \num{2e-2}$  	&	 $\num{2e-4}$  	&	0,15752444406	&	0,150171532079	&	[0.125343373890	;	0.211876034048]
\\ ${p_a}_{(0-24)}$ 	&	 proba asympt  	&	 $0.8 \pm \num{5e-2}$  	&	 $\num{5e-4}$  	&	0,898928642822	&	0,914677427502	&	[0.784148617203	;	0.952951629976]
\\ ${p_a}_{(25-44)}$ 	&	 proba asympt  	&	 $0.7 \pm \num{5e-2}$  	&	 $\num{5e-4}$  	&	0,829755760228	&	0,842949390508	&	[0.705750149416	;	0.905706359841]
\\ ${p_a}_{(45-64)}$ 	&	 proba asympt  	&	 $0.6 \pm \num{5e-2}$  	&	 $\num{5e-4}$  	&	0,720367229502	&	0,728090397493	&	[0.603294415455	;	0.811424601920]
\\ ${p_a}_{(65-74)}$ 	&	 proba asympt  	&	 $0.5 \pm \num{5e-2}$  	&	 $\num{5e-4}$  	&	0,547998162516	&	0,557910667894	&	[0.419668018619	;	0.647781943580]
\\ ${p_a}_{(75+)}$ 	&	 proba asympt  	&	 $0.4 \pm \num{5e-2}$  	&	 $\num{5e-4}$  	&	0,35917675405	&	0,352607578624	&	[0.239656937543	;	0.500570722355]
\\ ${p_a}_{h}$ 	&	 proba asympt  	&	 $0.3 \pm \num{5e-2}$  	&	 $\num{5e-4}$  	&	0,257884745616	&	0,257165852149	&	[0.125264553593	;	0.384904961687]
\\ $\delta_{(0-24)}$ 	&	 hospitalisation rate  	&	 $0.04 \pm \num{5e-3}$  	&	 $\num{5e-5}$  	&	0,010144659982	&	0,009600608688	&	[0.004439258498	;	0.017868355206]
\\ $\delta_{(25-44)}$ 	&	 hospitalisation rate  	&	 $0.045 \pm \num{5e-3}$  	&	 $\num{5e-5}$  	&	0,016232810879	&	0,01580132305	&	[0.007421965666	;	0.026313998135]
\\ $\delta_{(45-64)}$ 	&	 hospitalisation rate  	&	 $0.05 \pm \num{5e-3}$  	&	 $\num{5e-5}$  	&	0,024191029656	&	0,023879716645	&	[0.011679253926	;	0.039250445488]
\\ $\delta_{(65-74)}$ 	&	 hospitalisation rate  	&	 $0.055 \pm \num{5e-3}$  	&	 $\num{5e-5}$  	&	0,045459083851	&	0,045821702738	&	[0.033200901843	;	0.057723885149]
\\ $\delta_{(75+)}$ 	&	 hospitalisation rate  	&	 $0.06 \pm \num{5e-3}$  	&	 $\num{5e-5}$  	&	0,055505362466	&	0,055073221159	&	[0.045285162117	;	0.066858628289]
\\ $\delta_{h}$ 	&	 hospitalisation rate  	&	 $0.065 \pm \num{5e-3}$  	&	 $\num{5e-5}$  	&	0,064229352244	&	0,063849198302	&	[0.053489526728	;	0.077307620788]
\\ ${\gamma_a}_{(0-24)}$ 	&	 recover rate (asympt)  	&	 $0.29 \pm \num{2e-2}$  	&	 $\num{2e-4}$  	&	0,310519407671	&	0,302105514717	&	[0.261135695255	;	0.377880760318]
\\ ${\gamma_a}_{(25-44)}$ 	&	 recover rate (asympt)  	&	 $0.27 \pm \num{2e-2}$  	&	 $\num{2e-4}$  	&	0,265510561665	&	0,260660420927	&	[0.226038082064	;	0.321022777025]
\\ ${\gamma_a}_{(45-64)}$ 	&	 recover rate (asympt)  	&	 $0.25 \pm \num{2e-2}$  	&	 $\num{2e-4}$  	&	0,238445483855	&	0,233995795188	&	[0.199610989601	;	0.301033748724]
\\ ${\gamma_a}_{(65-74)}$ 	&	 recover rate (asympt)  	&	 $0.23 \pm \num{2e-2}$  	&	 $\num{2e-4}$  	&	0,20498707482	&	0,205516968503	&	[0.166480383656	;	0.241210524173]
\\ ${\gamma_a}_{(75+)}$ 	&	 recover rate (asympt)  	&	 $0.21 \pm \num{2e-2}$  	&	 $\num{2e-4}$  	&	0,157118841831	&	0,157207841006	&	[0.116686432030	;	0.200311085225]
\\ ${\gamma_a}_{h}$ 	&	 recover rate (asympt)  	&	 $0.19 \pm \num{2e-2}$  	&	 $\num{2e-4}$  	&	0,039249289793	&	0,038455690696	&	[0.016195067532	;	0.064961754286]
\\ ${\gamma_s}_{(0-24)}$ 	&	 recover rate (sympt)  	&	 $0.29 \pm \num{2e-2}$  	&	 $\num{2e-4}$  	&	0,327763131908	&	0,329919249227	&	[0.244533425017	;	0.399120340533]
\\ ${\gamma_s}_{(25-44)}$ 	&	 recover rate (sympt)  	&	 $0.27 \pm \num{2e-2}$  	&	 $\num{2e-4}$  	&	0,277338098622	&	0,277927876394	&	[0.215077058378	;	0.342582893244]
\\ ${\gamma_s}_{(45-64)}$ 	&	 recover rate (sympt)  	&	 $0.25 \pm \num{2e-2}$  	&	 $\num{2e-4}$  	&	0,2450803466	&	0,246465501586	&	[0.189853348410	;	0.303419146023]
\\ ${\gamma_s}_{(65-74)}$ 	&	 recover rate (sympt)  	&	 $0.23 \pm \num{2e-2}$  	&	 $\num{2e-4}$  	&	0,212419934074	&	0,209430480303	&	[0.172537773319	;	0.263014657947]
\\ ${\gamma_s}_{(75+)}$ 	&	 recover rate (sympt)  	&	 $0.21 \pm \num{2e-2}$  	&	 $\num{2e-4}$  	&	0,193171616525	&	0,191504351683	&	[0.159913850508	;	0.235828599167]
\\ ${\gamma_s}_{h}$ 	&	 recover rate (sympt)  	&	 $0.19 \pm \num{2e-2}$  	&	 $\num{2e-4}$  	&	0,17184608121	&	0,170755800091	&	[0.140590773429	;	0.204949330321]
\\ ${\gamma_q}_{(0-24)}$ 	&	 recover rate (hosp)  	&	 $0.07 \pm \num{5e-3}$  	&	 $\num{5e-5}$  	&	0,06066020319	&	0,059896730345	&	[0.050535982675	;	0.074505609344]
\\ ${\gamma_q}_{(25-44)}$ 	&	 recover rate (hosp)  	&	 $0.06 \pm \num{5e-3}$  	&	 $\num{5e-5}$  	&	0,05368037327	&	0,053478664903	&	[0.047685971046	;	0.060419067031]
\\ ${\gamma_q}_{(45-64)}$ 	&	 recover rate (hosp)  	&	 $0.05 \pm \num{5e-3}$  	&	 $\num{5e-5}$  	&	0,050090218261	&	0,049913064522	&	[0.045819342628	;	0.054874276919]
\\ ${\gamma_q}_{(65-74)}$ 	&	 recover rate (hosp)  	&	 $0.04 \pm \num{5e-3}$  	&	 $\num{5e-5}$  	&	0,048588797373	&	0,048331478227	&	[0.044849458139	;	0.052856080666]
\\ ${\gamma_q}_{(75+)}$ 	&	 recover rate (hosp)  	&	 $0.03 \pm \num{5e-3}$  	&	 $\num{5e-5}$  	&	0,047072146297	&	0,046903716917	&	[0.043455136800	;	0.051310737597]
\\ ${\gamma_q}_{h}$ 	&	 recover rate (hosp)  	&	 $0.02 \pm \num{5e-3}$  	&	 $\num{5e-5}$  	&	0,043961485856	&	0,044051929246	&	[0.038577846740	;	0.049056166152]
\\ $r_{(0-24)}$ 	&	 death rate  (hosp)  	&	 $0.01 \pm \num{2e-3}$  	&	 $\num{2e-5}$  	&	0,005198744739	&	0,005096628514	&	[0.003343925202	;	0.007274772619]
\\ $r_{(25-44)}$ 	&	 death rate  (hosp)  	&	 $0.015 \pm \num{2e-3}$  	&	 $\num{2e-5}$  	&	0,007489073117	&	0,007360831134	&	[0.005203349708	;	0.010126644743]
\\ $r_{(45-64)}$ 	&	 death rate  (hosp)  	&	 $0.02 \pm \num{2e-3}$  	&	 $\num{2e-5}$  	&	0,010912578086	&	0,010666367404	&	[0.007617534397	;	0.014996566858]
\\ $r_{(65-74)}$ 	&	 death rate  (hosp)  	&	 $0.025 \pm \num{2e-3}$  	&	 $\num{2e-5}$  	&	0,033576859624	&	0,03424532159	&	[0.022822230429	;	0.042526970102]
\\ $r_{(75+)}$ 	&	 death rate  (hosp)  	&	 $0.03 \pm \num{2e-3}$  	&	 $\num{2e-5}$  	&	0,040619671015	&	0,041035499863	&	[0.033566751223	;	0.047235364826]
\\ $r_{h}$ 	&	 death rate  (hosp)  	&	 $0.035 \pm \num{2e-3}$  	&	 $\num{2e-5}$  	&	0,050794267514	&	0,049220352816	&	[0.043703650451	;	0.063608031990]
\\ $\tilde r_h $ 	&	 death rate (homes)  	&	 $0.02 \pm \num{2e-3}$  	&	 $\num{2e-5}$  	&	0,061256871756	&	0,060835430597	&	[0.055271873752	;	0.069088303275]
\\ $P_\text{recovery}$ 	&	 care improvement  	&	 $0.7 \pm \num{5e-2}$  	&	 $\num{5e-4}$  	&	0,574824369393	&	0,581891705055	&	[0.493802053533	;	0.643503112810]
\\ $\mu_\text{recovery}$ 	&	 care improvement  	&	 $200 \pm \num{2e0}$  	&	 $\num{2e-2}$  	&	43,29766032639	&	41,18641719471	&	[33.39825429174	;	57.25690123013]
\\ $s_\text{recovery}$ 	&	 care improvement  	&	 $15 \pm \num{2e0}$  	&	 $\num{2e-2}$  	&	24,20494964973	&	23,98882599212	&	[18.96961120518	;	30.42122536558]
\\ SUPP$_{hosp}$ 	&	 supplementary entries  	&	 $1.15 \pm \num{5e-2}$  	&	 $\num{5e-4}$  	&	1,299991580674	&	1,299994214888	&	[1.299975318101	;	1.299999535025]
\\ $\mu_\text{hosp}$ 	&	 variable hosp. policy  	&	 $4000 \pm \num{5e2}$  	&	 $\num{5e0}$  	&	2385,31690003	&	2320,05566716	&	[1509.868785895	;	3290.433695411]
\\ $s_\text{hosp}$ 	&	 variable hosp. policy  	&	 $2000 \pm \num{2e2}$  	&	 $\num{2e0}$  	&	1549,6513557	&	1569,78395988	&	[1054.155873480	;	1958.642584287]
\\ $\text{delay}$ 	&	 variable hosp. policy  	&	 $15 \pm \num{2e1}$  	&	 $\num{2e-1}$  	&	10,66049481594	&	10,60548108661	&	[8.482922969255	;	12.88913656094]
\\ $P_{cor} $ 	&	 COVID-19 related deaths  	&	 $0.8 \pm \num{5e-2}$  	&	 $\num{5e-4}$  	&	0,811351608526	&	0,831187765137	&	[0.679571638618	;	0.893421854522]
\\ $P_{th}$ 	&	 transmission to homes  	&	 $10 \pm \num{1e0}$  	&	 $\num{1e-2}$  	&	23,53680834660	&	23,54512032772	&	[19.93376338146	;	27.02338028570]
\\ $P^\prime_{th}$ 	&	 transmission to homes  	&	 $10 \pm \num{1e0}$  	&	 $\num{1e-2}$  	&	20,79289318079	&	20,79659960089	&	[17.28671758114	;	23.60964091171]
\\ $m_h$ 	&	 transmission in homes  	&	 $0.5 \pm \num{5e-2}$  	&	 $\num{5e-4}$  	&	0,158681848021	&	0,149777130915	&	[0.019275156418	;	0.325822901192]
\\ $C_\text{homelock}$ 	&	 contacts coefficient  	&	 $0.5 \pm \num{1e-2}$  	&	 $\num{1e-4}$  	&	0,44237123715	&	0,439827890287	&	[0.403292590770	;	0.483764155961]
\\ $C_\text{worklock}$ 	&	 contacts coefficient  	&	 $0.1 \pm \num{1e-2}$  	&	 $\num{1e-4}$  	&	0,051829310886	&	0,049917692181	&	[0.017249627277	;	0.093200331023]
\\ $C_\text{leisurelock}$ 	&	 contacts coefficient  	&	 $0.1 \pm \num{1e-2}$  	&	 $\num{1e-4}$  	&	0,084741835898	&	0,085008995606	&	[0.055234313503	;	0.116113053110]
\\ $C_\text{homeunlock}$ 	&	 contacts coefficient  	&	 $0.55 \pm \num{1e-2}$  	&	 $\num{1e-4}$  	&	0,511199902073	&	0,513939143457	&	[0.469304521115	;	0.544015515231]
\\ $C_\text{workunlock}$ 	&	 contacts coefficient  	&	 $0.15 \pm \num{1e-2}$  	&	 $\num{1e-4}$  	&	0,096156684452	&	0,093226266898	&	[0.060220232170	;	0.142714061020]
\\ $C_\text{schoolunlock}$ 	&	 contacts coefficient  	&	 $0.15 \pm \num{1e-2}$  	&	 $\num{1e-4}$  	&	0,233790909687	&	0,234328214482	&	[0.204769173704	;	0.262534237400]
\\ $C_\text{leisurejune}$ 	&	 contacts coefficient  	&	 $0.15 \pm \num{1e-2}$  	&	 $\num{1e-4}$  	&	0,153612935119	&	0,155637045541	&	[0.099643699322	;	0.201904481695]
\\ $C_\text{leisurejuly}$ 	&	 contacts coefficient  	&	 $0.3 \pm \num{1e-2}$  	&	 $\num{1e-4}$  	&	0,442756909838	&	0,438529235296	&	[0.384476561231	;	0.507646484990]
\\ $C_\text{leisureaug}$ 	&	 contacts coefficient  	&	 $0.2 \pm \num{1e-2}$  	&	 $\num{1e-4}$  	&	0,062426540365	&	0,057915018224	&	[0.003577650385	;	0.137566816909]
\\ $C_\text{schoolsept}$ 	&	 contacts coefficient  	&	 $0.2 \pm \num{4e-2}$  	&	 $\num{5e-4}$  	&	0,825882503926	&	0,88168229348	&	[0.405336998472	;	0.989817635232]
\\ $C_\text{leisuresept}$ 	&	 contacts coefficient  	&	 $0.25 \pm \num{4e-2}$  	&	 $\num{5e-4}$  	&	0,334161870284	&	0,312632099794	&	[0.212125482364	;	0.555907162072]
\\ $C_\text{reimp}$ 	&	 reimportation coefficient  	&	 $50 \pm \num{5e0}$  	&	 $\num{5e-2}$  	&	29,46476967439	&	30,04056589424	&	[17.06883610257	;	40.46199597591]
\\ \hline
\end{tabular}
}
\caption{Complete list of estimated parameters from October 31, 2020 data}
\label{list_par}
\end{table}

\end{appendix}

\end{document}